\documentclass[pre,showpacs,twocolumn,preprintnumbers,amsmath,amsfonts,amssymb,floatfix,aps,superscriptaddress]{revtex4}

\usepackage{graphicx}
\usepackage{epsfig}
\usepackage[normalem]{ulem}
\usepackage{color}
\usepackage{hyperref}

\graphicspath{{./fig-jpg/}{./fig-ps/}{./figures/}{./Figs/}}

\definecolor{mygreen}{rgb}{0,0.7,0}

\begin{document}
\title{Nonlinear Waves in an Experimentally Motivated Ring-shaped Bose-Einstein Condensate Setup}

\date{\today}

\author{M. Haberichter}%
\email{mareike@math.umass.edu}
\author{P. G. Kevrekidis}%
\affiliation{%
Department of Mathematics and Statistics, University of Massachusetts, Amherst, Massachusetts 01003-4515, USA
}%
\author{R. Carretero-Gonz{\'a}lez}
\affiliation{Nonlinear Dynamical Systems
Group,\footnote{\texttt{URL}: http://nlds.sdsu.edu}
Computational Sciences Research Center, and
Department of Mathematics and Statistics,
San Diego State University, San Diego, California 92182-7720, USA}
\author{M. Edwards}
\email{edwards@georgiasouthern.edu}
\affiliation{
Department of Physics, Georgia Southern University, 
Statesboro, Georgia 30460-8031, USA}
\affiliation{
Joint Quantum Institute, National Institute of Standards and 
Technology and the University of Maryland,
Gaithersburg, Maryland 20899, USA}

\begin{abstract}
We systematically construct stationary soliton states in a one-component, 
two-dimensional, repulsive, Gross-Pitaevskii equation with a ring-shaped
target-like trap similar to the potential used to confine a Bose-Einstein 
condensate in a recent experiment [Eckel, et al.\ {\em Nature} {\bf 506}, 200 (2014)].
In addition to the ground state configuration, we identify a
wide variety of excited states involving phase jumps (and associated
dark solitons) inside the ring.
These configurations are obtained from a systematic bifurcation
analysis starting from the linear, small 
atom density, limit.
We study the stability, and when unstable, the dynamics of the most
basic configurations. Often these lead to vortical dynamics
inside the ring persisting over long time scales in our numerical
experiments.
To illustrate the relevance of the identified
states, we showcase how such dark-soliton
configurations (even the unstable ones)
can be created in laboratory condensates by using
phase-imprinting techniques.
\end{abstract}
\maketitle

%%%%%%%%%%%%%%%%%%%%%%%%%%%%%%%%%%%%%%%%%%%%%%%%%%%%%%%%%%%%%%%%%%%%%%%%%%%%%%
\section{Introduction}
\label{Intro}
%%%%%%%%%%%%%%%%%%%%%%%%%%%%%%%%%%%%%%%%%%%%%%%%%%%%%%%%%%%%%%%%%%%%%%%%%%%%%%

Atomic Bose-Einstein condensates (BECs) \cite{pethick_smith_2008,BEC_Stringari,EmergentNL,LA-UR-15-20791} offer an ideal testing ground for confronting theoretical models of nonlinear matter waves with experimental data. Since their
experimental realization, there have been tremendous advances \cite{LA-UR-15-20791,RevModPhys.74.875,RevModPhys.71.463,RevModPhys.74.1131,PhysRevLett.89.040401} in trapping, guiding, manipulating and controlling BECs. For instance, recent advances in all-optical trapping~\cite{1367-2630-11-4-043030,dig_hol,Pasienski:08} have produced confined atomic clouds with  temperatures at the nanokelvin scale. All-optical trapping, in turn, has enabled the strength of the atom-atom interactions in atomic gas BECs to be tuned to any desired value over many orders of magnitude~\cite{rhh} by adjusting an external magnetic field through the phenomenon of the Feshbach resonance~\cite{Inouye1998Observation}. This enables a wide range of experiments to be conducted because the properties of BECs ---as well as the nature of their effective
  nonlinearity--- crucially depend on the strength and sign of these interactions. 

These advances have led to more stable, easier to use experimental settings and high-precision measurements of coherent structures in BECs. In a plethora of experiments, matter-wave dark~\cite{2010JPhA.43u3001F} and
bright~\cite{rhh2,tomio,Khaykovich1290,Strecker2002,PhysRevLett.96.170401} solitons have been realized in single- and multi-component BECs  with repulsive or attractive interatomic interactions, respectively. For example, bright solitons have been formed in ultracold $^{7}$Li gas~\cite{Strecker2002,Khaykovich1290} as well as during the collapse of $^{85}$Rb condensates~\cite{PhysRevLett.96.170401}. Dark solitons have been studied in $^{87}$Rb condensates \cite{PhysRevLett.83.5198,PhysRevLett.86.2926,2008NatPh.4.496B,markus} and in sodium BECs \cite{Denschlag97,Dutton663}. Furthermore, coupled dark-bright  solitons have been engineered in $^{87}$Rb condensates using phase-imprinting methods \cite{2008NatPh.4.496B} or generated during superfluid-superfluid counterflow \cite{PhysRevLett.106.065302,MIDDELKAMP2011642}. Finally, matter wave gap solitons \cite{gap,RevModPhys.78.179} have been produced in BECs trapped in light-induced periodic potentials.

At the theoretical level, and for sufficiently low temperatures,
static and dynamical properties of BECs have been quite successfully
modeled by an effective mean-field equation known as the Gross-Pitaevskii 
equation (GPE)~\cite{pethick_smith_2008,BEC_Stringari,doi:10.1137/1.9781611973945}. The GPE is tantamount to a (cubic) nonlinear Schr\"odinger (NLS)
equation with the addition of the external potential that confines the BEC.
The $(2+1)$-dimensional version of the fully 3D equation reads, in terms of physical 
units, as
\begin{eqnarray}
i\hbar\partial_t\Phi=\Big[-\frac{\hbar^2}{2m}\nabla^2+g_{\rm 2D}|\Phi|^2+V(\boldsymbol{r})\Big]\Phi\,,
\label{NLS_units}
\end{eqnarray}
where $\Phi(\boldsymbol{r},t)$ is the macroscopic BEC wavefunction, 
$\nabla^2$ is the Laplacian in $\boldsymbol{r}=(x,y)$, $m$ is the atomic mass and 
$g_{\rm 2D}$ describes the effective 2D strength of the atom-atom interaction. 
The effective 2D coupling constant $g_{\rm 2D}$ is given by 
$g_{\rm 2D}=g/(\sqrt{2\pi}a_z)=2\sqrt{2\pi}\hbar a_z\omega_z a$,
where $\omega_z$ is the harmonic trapping strength in the transverse direction, 
with $a_z$ being its corresponding harmonic oscillator length. 
The 3D coupling constant is $g=4\pi\hbar^2a/m$, where $a$ is the $s$-wave scattering length.

In the following, we set $g_{\rm 2D}>0$, that is the nonlinearity in the GPE is chosen to be defocusing \cite{doi:10.1137/1.9781611973945,2004dcns.bookA,sulem2007nonlinear} which models a repulsive interatomic interaction, as is the
case, e.g., in $^{87}$Rb. Multiple stationary dark-soliton states can emerge when the repulsion between dark solitons is counterbalanced by the inclusion of a trapping potential $V(\boldsymbol{r})$ in Eq.~(\ref{NLS_units}). The existence and formation of nonlinear patterns in BECs crucially depend on the chosen form for the applied trapping potential $V(\boldsymbol{r})$. The traditionally used magnetic traps can be adequately modeled by an harmonic external potential of the form \cite{RevModPhys.71.463,doi:10.1142/S0217984904006809}
\begin{eqnarray}\label{V_para}
V=\frac{1}{2}m\left(\omega_x^2 x^2+\omega_y^2 y^2\right)\,,
\end{eqnarray}
where, for generality, the trap frequencies $\omega_x$ and $\omega_y$ along the $x$- and $y$-direction can be chosen to be different. Static and dynamical properties of matter-wave dark solitons have been investigated in great detail in model (\ref{NLS_units}) with the parabolic confining potential (\ref{V_para}) and higher-dimensional analogues thereof. For example, dark soliton stripes and multivortex states such as vortex dipoles, tripoles, and quadrupoles have been found \cite{PhysRevA.71.033626,PhysRevA.74.023603} and their existence, stability and dynamics have been discussed in detail in the literature \cite{PhysRevA.82.013646,MIDDELKAMP20111449}.  

However, in recent years there has been increasing research activity in exploring 
different choices (specifically non-parabolic ones) for the external trapping 
potential in Eq.~(\ref{NLS_units}). Examples of trapping configurations recently used in BEC experiments include: double \cite{PhysRevA.55.4318,PhysRevA.59.1488,PhysRevA.64.011601,PhysRevA.72.021604,PhysRevLett.94.090405,PhysRevLett.95.010402,tilman}, and more-well (such as four-well~\cite{2009PhRvE.80d6611W})
potentials, box potentials \cite{BEC_Stringari}, optical lattice potentials \cite{BEC_Stringari,ADHIKARI2003229,PhysRevLett.82.2022}, or magnetic quadrupole trap combined with an optical dipole trap \cite{PhysRevA.79.063631}, among
many others.

In this article, we wish to explore the existence and stability of localized states in the two-dimensional (2D) GPE (\ref{NLS_units}) with a ring-shaped trapping potential and repulsive interatomic interactions.
A key feature of our work is the identification of a wide variety of
nonlinear states in this system including ones bearing different
numbers of phase jumps and associated dark solitons.  The
bifurcation analysis of such stationary solutions is complemented
by the corresponding stability analysis, and the dynamical evolution
of potentially unstable configurations. Equally importantly, 
phase imprinting protocols are utilized in suitably crafted
numerical experiments in
order to illustrate the potential of such states towards being
realized in recently considered experimental setups.

More specifically, our considerations are tailored
the recent experimental setup of atomtronic systems \cite{PhysRevLett.103.140405,PhysRevA.75.023615}, that are confined, neutral, ultracold atomic gases which exhibit behavior analogous to semiconductor electronic devices and circuits. 
In atomtronics, ring BECs are used \cite{PhysRevLett.106.130401,PhysRevLett.110.025302,PhysRevA.88.053615} to realize atomic-gas analogs of superconducting quantum interference devices (SQUIDs). In Ref.~\cite{PhysRevLett.106.130401}, a  closed-loop atom circuit was implemented for the first time in a ring-shaped confining potential. Rf SQUIDs \cite{BraginskiBook} have been created \cite{PhysRevLett.110.025302} in ring BECs by rotating a weak link (a localized region of reduced superfluid density) around the ring-shaped condensate. 
A rotating weak link was used to drive phase slips which changed the circulation 
around the ring and simulations, based on the GPE, showed how the circulation of 
the ring BEC can be probed by measuring the distribution of hole areas in 
time-of-flight images~\cite{PhysRevA.88.053615}.
We also note in passing that ring-shaped BECs have been
recently argued~\cite{Eckel:2017uqx} as an interesting laboratory testbed
for cosmological physics.

The article is structured as follows. In Sec.~\ref{Model} we briefly review some of the properties of the GPE in $(2+1)$ dimensions  and introduce the chosen ring-shaped trapping potential. For a detailed discussion of the existence and stability analysis of steady-state solutions in the 2D GPE with repulsive interactions we refer the interested reader to the reviews and textbooks \cite{doi:10.1137/1.9781611973945,2004dcns.bookA,EmergentNL}.  Our numerical results are reported in Sec.~\ref{Num}. Finally, in Sec.~\ref{Con}, we summarize our conclusions and discuss possible directions for further work.

%%%%%%%%%%%%%%%%%%%%%%%%%%%%%%%%%%%%%%%%%%%%%%%%%%%%%%%%%%%%%%%%%%%%%%%%%%%%%%
\section{Model and Methodology}
\label{Model}
%%%%%%%%%%%%%%%%%%%%%%%%%%%%%%%%%%%%%%%%%%%%%%%%%%%%%%%%%%%%%%%%%%%%%%%%%%%%%%

To simplify our numerical calculations, we rewrite Eq.~(\ref{NLS_units}) in 
its well-known dimensionless form~\cite{doi:10.1137/1.9781611973945,EmergentNL}
\begin{eqnarray}
i\partial_t\Phi=-\frac{1}{2}\nabla^2\Phi+|\Phi|^2\Phi+V(\boldsymbol{r})\Phi\,,
\label{NLS_eq}
\end{eqnarray}
where $\Phi=\Phi(x,y)$ is the 2D wavefunction and $\nabla^2$ 
is the Laplacian in $\boldsymbol{r}=(x,y)$. Equation~(\ref{NLS_eq}) is obtained from 
Eq.~(\ref{NLS_units}) by averaging (integrating) along the $z$-direction and
rescaling space coordinates by the the transverse oscillator length $a_z$ 
and time by $\omega_z^{-1}$. Then, the density $|\Phi|^2$, length, 
time and energy are respectively measured in units of $(2\sqrt{2\pi}aa_z)^{-1}$, the 
harmonic oscillator length $a_z=\sqrt{\hbar/(m\omega_z)}$, the inverse trap frequency 
$\omega_z^{-1}$ and energy $\hbar\omega_z$. 

%%%%%%%%%%%%%%%%%%%%%%%%%%%%%%%%%%%%%%%%%%%%%%%%%%%%%%%%%%%%%%%%%%%%%%%%%%%
\begin{figure}[!htb]
\includegraphics[width=6cm]{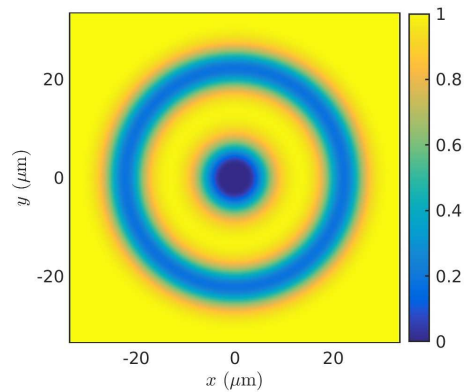}
\caption{(Color online)
Ring-shaped trapping potential $V$, given in Eq.~(\ref{Vtrap}), corresponding
to an experiment performed at NIST~\cite{PhysRevA.92.033602}.
In this figure, and all subsequent ones, space $(x,y)$ is displayed
using physical units (in microns).
}
\label{Trap_Pot}
\end{figure}
%%%%%%%%%%%%%%%%%%%%%%%%%%%%%%%%%%%%%%%%%%%%%%%%%%%%%%%%%%%%%%%%%%%%%%%%%%%

We choose an external trapping potential as experimentally obtained from a fit provided by NIST experimentalists corresponding to a ring-shaped channel of mean radius $r_\text{ring}$ together with a central well of radius $r_\text{disk}$. Stationary ground-state condensates filling this potential (see Fig.\ \ref{Trap_Pot}) consist of a central disk surrounded by a ring thus motivating the names $r_\text{disk}$ and $r_\text{ring}$~\cite{PhysRevA.92.033602,eckel_2013}.
This potential has the flexibility to be either a ring-plus-disk or just a ring.  In the case where a ring is present, the disk can be used as a phase reference to detect phase variations in the ring caused by, e.g., stirring.
Specifically, the fitted potential from the experiments takes the radial form:
\begin{eqnarray}
V(r)=\begin{cases} 
      1-A\,e^{-\frac{(r-r_{\text{ring}})^2}{s_{\text{ring}}^2}}
      -e^{-\frac{(r-r_{\text{disk}})^2}{s_{\text{disk}}^2}} 
& r\ge r_{\text{disk}} 
\\[2.0ex]
      -A\, e^{-\frac{(r-r_{\text{ring}})^2}{s_{\text{ring}}^2}} & r < r_{\text{disk}} \,,
   \end{cases}
\label{Vtrap}   
\end{eqnarray}
where $r_{\text{ring}}$, $A$ and $s_{\text{ring}}$ represent, respectively, the radius, 
the amplitude and the width of this ring-shaped potential. 
The experimentally fitted potential parameters correspond to:
$r_\text{ring} = 22.27$ $\mu$m, 
$r_\text{disk} = 2.597$ $\mu$m,  
$s_\text{ring} = 3.913$ $\mu$m, 
$s_\text{disk} = 4.717$ $\mu$m, and 
$A  = 0.8206$. 
Expressed in terms of the dimensionless units of Eq.~(\ref{NLS_eq}), based on a 
transverse trap frequency $\omega_z/2\pi = 500$ Hz, these quantities correspond to:
$r_{\text{ring}}=25.304738$, 
$r_{\text{disk}}=2.95089$, 
$s_{\text{ring}}=4.446226$,
$s_{\text{disk}}=5.3597867$, and 
$A  = 0.8206$. 
A plot of the resulting ring-shaped potential is displayed in Fig.~\ref{Trap_Pot}.
Note that for ease of interpretation, we opt to display in this figure, and all 
subsequent ones, the spatial dimensions in the original variables, namely in microns.

Let us now construct stationary solutions of Eq.~(\ref{NLS_eq}) by separating space
and time according to
\begin{eqnarray}
\Phi(\boldsymbol{r},t)=\phi(\boldsymbol{r})e^{-i\mu t}\,,
\label{phi_ansatz}
\end{eqnarray}
where $\mu$ is the (dimensionless) chemical potential. Substituting ansatz~(\ref{phi_ansatz}) 
into the 2D GPE~(\ref{NLS_eq}) yields the steady-state equation
\begin{eqnarray}\label{steady}
-\frac{1}{2}\nabla^2\phi+|\phi|^2\phi+\left[V(x,y)-\mu\right]\phi=0.
\end{eqnarray}
Steady-state solutions for Eq.~(\ref{steady}) correspond to mono-parametric branches
parametrized by the chemical potential $\mu$ which, in turn, fixes the number
of BEC atoms in the condensate. This relationship is obtained through the 
conserved quantity of the GPE corresponding to the (squared)
$L^2$ norm of the solution:
\begin{eqnarray}
\label{Dens}
N=\iint_{-\infty}^{+\infty} \left|\phi(x,y)\right|^2\, dx\,dy.
\end{eqnarray}
Thus, after bringing back the dimensions into Eq.~(\ref{Dens}), $N$ can be 
identified with the mass or total number of atoms in the BEC.
In what follows we find suitable starting points on a given solution branch
and then vary $\mu$ using continuation methods to follow the entire branch
possibly leading to bifurcations (when two solution branches collide or when 
new branches emanate from existing ones) as the chemical potential $\mu$ 
is varied~\cite{CHARALAMPIDIS2018482,MIDDELKAMP20111449}. 
For given chemical potential $\mu$, we find stationary nonlinear solutions to 
Eq.~(\ref{steady}) by using two different implementations of Newton algorithms. 
Details on these numerical methods are found in Sec.~\ref{Num}.

After having numerically computed solutions, for each chosen value of $\mu$, we 
proceed to study their instability modes by performing the well-known Bogoliubov-de 
Gennes (BdG) stability analysis~\cite{BEC_Stringari,pethick_smith_2008,EmergentNL}. 
We perturb around a stationary solution $\phi_0$ using the perturbation ansatz
\begin{eqnarray}\label{BdG}
\phi(\boldsymbol{r})=\phi_0(\boldsymbol{r})+\left[a(\boldsymbol{r})e^{i\omega t}+b^\star(\boldsymbol{r})e^{-i\omega^\star t}\right]\,,
\end{eqnarray}
where $(\cdot)^\star$ denotes complex conjugation and $\omega$ is a complex eigenfrequency. 
Linearization of the GPE~(\ref{NLS_eq}) around the stationary solution $\phi_0$ via the 
ansatz (\ref{BdG}) yields the following BdG eigenvalue problem
\begin{eqnarray}\label{BdGEVproblem}
-\omega \begin{pmatrix} 
a \\
b 
\end{pmatrix} =\begin{pmatrix} 
A_{11} & A_{12} \\
-A_{12}^\star & -A_{11} 
\end{pmatrix} \begin{pmatrix} 
a \\
b
\end{pmatrix}\,,
\end{eqnarray}
where the matrix elements are explicitly given by 
\begin{subequations}
\begin{align}\label{BdGmatrixelements}
A_{11}&=-\frac{1}{2}\nabla^2+2|\phi_0|^2+V(x,y)-\mu\,.\\
A_{12}&=\left(\phi_0\right)^2\,.
\end{align}
\end{subequations}
We compute the eigenfunctions $\{a(x,y),b(x,y)\}$ and eigenfrequencies $\omega$ of the BdG 
eigenvalue problem (\ref{BdGEVproblem}) for a steady-state solution $\phi_0$ and for a
given value $\mu$ using the $\tt{eigs}$ MATLAB 
routine~\cite{doi:10.1137/S0895479800371529,doi:10.1137/1.9780898719628} and our results 
are further checked with the Scalable Library for Eigenvalue Problem Computations 
(SLEPc)~\cite{Hernandez:2005:SSF,Hernandez:2003:SSL,slepc-users-manual}. 
The BdG stability results are then depicted in terms of the corresponding spectra by 
plotting the real and imaginary parts of the eigenfrequencies as a function of $\mu$. 
Recall that for a linearly (neutrally) stable soliton configuration, all eigenfrequencies 
must be real, that is $\text{Im}(\omega)=0$.

%%%%%%%%%%%%%%%%%%%%%%%%%%%%%%%%%%%%%%%%%%%%%%%%%%%%%%%%%%%%%%%%%%%%%%%%%%%%%%
\section{Numerical Simulations}
\label{Num}
%%%%%%%%%%%%%%%%%%%%%%%%%%%%%%%%%%%%%%%%%%%%%%%%%%%%%%%%%%%%%%%%%%%%%%%%%%%%%%

Our numerical results are based on discretizing the ensuing nonlinear equations
---for the dynamics Eq.~(\ref{NLS_eq}), for the steady states Eq.~(\ref{steady}), and
for the BdG spectra eigenvalue problem Eq.~(\ref{BdGEVproblem})--- on the rectangular, 
uniform, 2D grid $(x,y)\in[-50,50]$ and with grid spacing $\Delta x=0.2$.
% that is we choose $501$ grid points in $x$- and $y$-direction. 
%
The steady-state equation~(\ref{steady}) is solved using a Newton-Krylov 
algorithm~\cite{Kelly} and then, the obtained states are checked using Newton 
iterations implemented in the {\tt{SNES}} libraries of 
{\tt{PETSc}}~\cite{petsc-web-page,petsc-user-ref,petsc-efficient}. 

In order to pick a suitable initial guess for convergence towards the steady
state we use the first few solutions close to the linear limit. The linear limit,
corresponding to weak nonlinearities in Eq.~(\ref{BdGEVproblem}), may be
formally identified with $N\rightarrow 0$. 
Then, stationary states for larger values of $\mu$ are obtained via numerical 
continuation by taking as initial guess the configuration calculated at nearby 
chemical potential values. 
The numerical results presented below were carried out with the chemical potential 
$\mu$ varying over the interval $[0,1]$ with steps of $\Delta \mu=0.002$. 
If not otherwise stated, all configurations depicted here correspond to the
chemical potential $\mu=0.9$.

Further insights into the dynamical properties and stability of the found 
steady states can be obtained by perturbing these solutions with the eigenvectors,
computed in the BdG linearization analysis (\ref{BdGEVproblem}), and studying 
their temporal evolution. To simulate the time evolution based on Eq.~(\ref{NLS_eq}), 
we employ a fourth order Runge-Kutta integrator in time with a second-order finite 
differences used for the discretization of the spatial derivatives. 
%

%%%%%%%%%%%%%%%%%%%%%%%%%%%%%%%%%%%%%%%%%%%%%%%%%%%%%%%%%%%%%%%%%%%%%%%%%%%
\begin{figure}[t]
\includegraphics[width=\columnwidth]{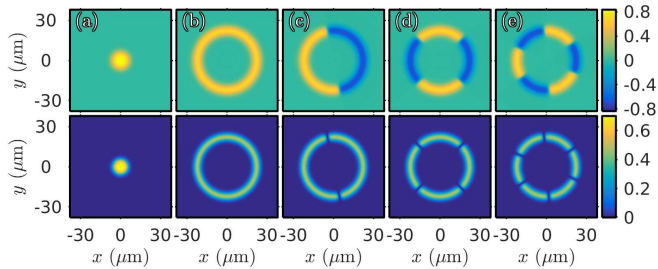}
\caption{(Color online)
Ground state and $n$-dark soliton solutions for $\mu=0.9$.
The real part and density of the solutions are depicted, respectively,
in the top and bottom rows of panels.
(a) Ground state (that populates the central well of the external potential). 
(b) Basic ring state without any dark solitons.
(c)--(e) First 3 excited states along the ring containing, respectively, 
two, four, and six dark solitons.
All these stationary solutions are purely real.
}
\label{Soli_Funda}
\end{figure}
%%%%%%%%%%%%%%%%%%%%%%%%%%%%%%%%%%%%%%%%%%%%%%%%%%%%%%%%%%%%%%%%%%%%%%%%%%%

%%%%%%%%%%%%%%%%%%%%%%%%%%%%%%%%%%%%%%%%%%%%%%%%%%%%%%%%%%%%%%%%%%%%%%%%%%%
\subsection{States bifurcating from the linear limit}
%%%%%%%%%%%%%%%%%%%%%%%%%%%%%%%%%%%%%%%%%%%%%%%%%%%%%%%%%%%%%%%%%%%%%%%%%%%

\begin{figure}
\centering
\includegraphics[width=\columnwidth]{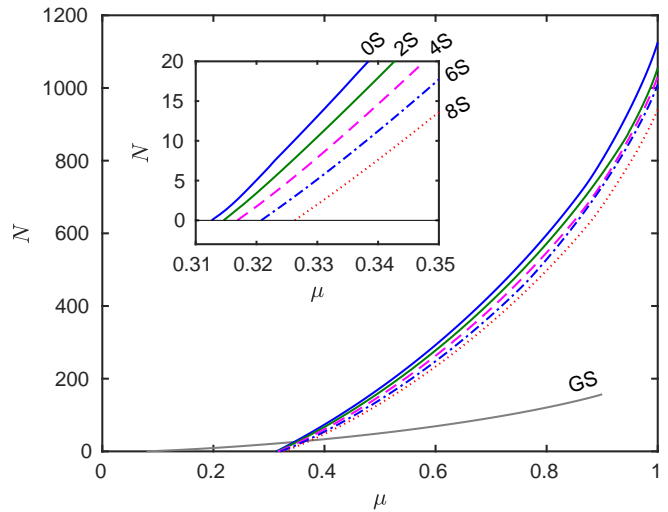}
\caption{(Color online)
Particle number $N$ as a function of $\mu$ for the ground state (GS) and the
$n$-soliton ($n$S) stationary steady states. These steady states are obtained
by continuation from the $N\simeq 0$ limit where the solutions are calculated 
by taking an initial guess in our fixed point iterations corresponding to the 
first few eigenfunctions (excited states) in the linear limit.
The critical chemical potential values $\mu_{\text{crit}}$ at which the
different states are found to emerge correspond to
$\mu_{\text{crit}}^{(0)}=0.313$ for 0S,
$\mu_{\text{crit}}^{(2)}=0.314$ for 2S,
$\mu_{\text{crit}}^{(4)}=0.316$ for 4S,
$\mu_{\text{crit}}^{(6)}=0.320$ for 6S, and
$\mu_{\text{crit}}^{(8)}=0.326$ for 8S.
The corresponding profiles for these solutions for $\mu=0.9$ are
depicted in Fig.~\ref{Soli_Funda}.
}
\label{Number_atoms_Fund}
\end{figure}

The most basic steady state is given by the ground state. For our
system with the potential given in Eq.~(\ref{Vtrap}), the ground state
emerging from the linear limit simply corresponds to a localized ``hump''
of atoms that populate the central well of the potential (see 
panels (a) in Fig.~\ref{Soli_Funda}). The corresponding
particle number (or mass) for the ground state branch as a function of the
chemical potential is depicted in Fig.~\ref{Number_atoms_Fund} (see line
denoted by GS). It is interesting to note that the ground state does not 
populate the ring of the external potential. 
In fact the ring does not get populated until  $\mu$ reaches 
$\mu\simeq \mu_{\text{crit}}^{(0)}=0.313$ 

For $\mu\geq \mu_{\text{crit}}^{(0)}$ a new state emerges from the linear, $N\simeq0$, limit that starts filling the ring with atoms (see panels (b) in 
Fig.~\ref{Soli_Funda}). This ring-shaped solution would correspond to the
ground state if the central well was absent. The mass for this
ring state is depicted in Fig.~\ref{Number_atoms_Fund} (see line denoted by 0S).
Since this ring state could be considered as a quasi-1D periodic line of density,
it is possible to think about the configurations stemming from its excited states.

For instance, in an infinite 1D line density, in the absence of external potential, the repulsive GPE admits a dark soliton solution~\cite{EmergentNL,doi:10.1137/1.9781611973945}
corresponding to the first excited state.
In the case of the ring line density, the wavefunction necessarily has to be periodic 
along the ring. This topological constraint restricts the number of dark solitons
that can be excited along the ring to be an {\em even} number. With an even
number of dark solitons along the ring, periodic boundaries are automatically
satisfied.

We show these $n$-soliton steady-state solutions in Fig.~\ref{Soli_Funda} for $n=0$ 
(the ring state without any solitons), $n=2$ (a pair of dark solitons), $n=4$
(two dark soliton pairs), and $n=6$ (three dark soliton pairs). Note that, due
to symmetry, in the steady state all the dark solitons must be equidistant from each 
other along the periodic ring.
The particle numbers corresponding to these $n$-soliton solutions are 
depicted in Fig.~\ref{Number_atoms_Fund}.
Note that the $n$-dark soliton solutions, populating the ring, bifurcate from 
the linear limit ($N\simeq 0$) and are independent of the ground state that 
populates the central well.

%%%%%%%%%%%%%%%%%%%%%%%%%%%%%%%%%%%%%%%%%%%%%%%%%%%%%%%%%%%%%%%%%%%%%%%%%%%
%%%%%%%%%%%%%%%%%%%%%%%%%%%%%%%%%%%%%%%%%%%%%%%%%%%%%%%%%%%%%%%%%%%%%%%%%%%

%%%%%%%%%%%%%%%%%%%%%%%%%%%%%%%%%%%%%%%%%%%%%%%%%%%%%%%%%%%%%%%%%%%%%%%%%%%
\begin{figure}[!htb]
\includegraphics[width=0.95\columnwidth]{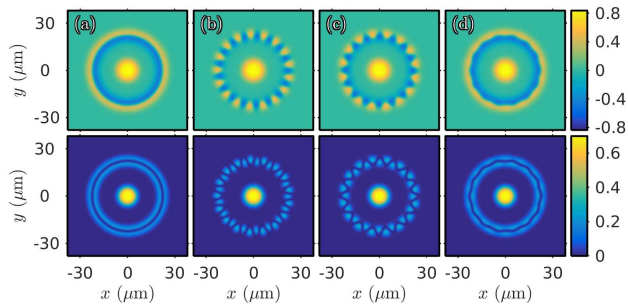}
\caption{(Color online)
Double-ring solution and some of its bifurcating states for $\mu=0.7$.
(a) Double ring solution (that bifurcated
from the ground state) consisting of two concentric out-of-phase rings.
(b)--(d) Successive
states bifurcating away from the double-ring solution.
The corresponding particle numbers for these solutions as a function
of $\mu$ are depicted in Fig.~\ref{Number_atoms_Ring}.
Same layout as in Fig.~\ref{Soli_Funda}.
}
\label{Ring_Exci}
\end{figure}
%%%%%%%%%%%%%%%%%%%%%%%%%%%%%%%%%%%%%%%%%%%%%%%%%%%%%%%%%%%%%%%%%%%%%%%%%%%

%%%%%%%%%%%%%%%%%%%%%%%%%%%%%%%%%%%%%%%%%%%%%%%%%%%%%%%%%%%%%%%%%%%%%%%%%%%
\begin{figure}
\centering
\includegraphics[width=0.95\columnwidth]{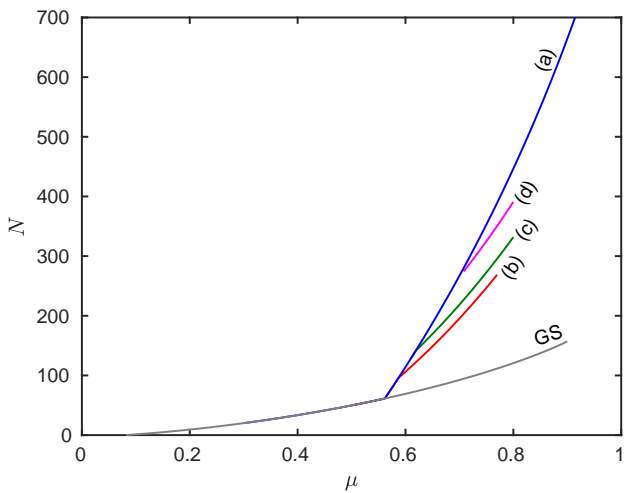}
\caption{(Color online)
Particle number $N$ as a function of $\mu$ for ground state (GS), 
the double-ring (a) and its first three bifurcating branches (b)--(d). 
The corresponding profiles for $\mu=0.7$ are depicted in Fig.~\ref{Ring_Exci}.
These double-ring bifurcates from the ground state at $\mu\simeq0.560$,
while the subsequent states bifurcate in turn from the double-ring solution
for (b) $\mu=0.586$, (c) 0.618, and (d) 0.708.
}
\label{Number_atoms_Ring}
\end{figure}
%%%%%%%%%%%%%%%%%%%%%%%%%%%%%%%%%%%%%%%%%%%%%%%%%%%%%%%%%%%%%%%%%%%%%%%%%%%

%%%%%%%%%%%%%%%%%%%%%%%%%%%%%%%%%%%%%%%%%%%%%%%%%%%%%%%%%%%%%%%%%%%%%%%%%%%
\begin{figure}[!htb]
\includegraphics[width=1.00\columnwidth]{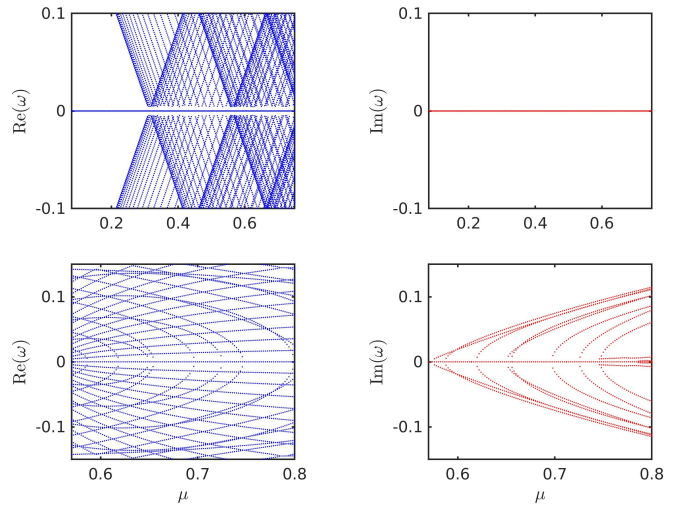}
\caption{(Color online)
Stability BdG spectra for the ground state (top row of panels) and the double-ring state
(bottom row of panels) as a function of the chemical potential $\mu$. The corresponding 
profiles are depicted in the first row of panels of, respectively, Fig.~\ref{Soli_Funda} 
and Fig.~\ref{Ring_Exci}.
The left and right panel depict, respectively, the real and imaginary parts
of the spectra. Recall that (neutral) stability is only achieved when
Im$(\omega)=0$.
The ground state is always (neutrally) stable while the double-ring state
is, since its inception, always unstable.
}
\label{Gr_Spectra}
\end{figure}
%%%%%%%%%%%%%%%%%%%%%%%%%%%%%%%%%%%%%%%%%%%%%%%%%%%%%%%%%%%%%%%%%%%%%%%%%%%

%%%%%%%%%%%%%%%%%%%%%%%%%%%%%%%%%%%%%%%%%%%%%%%%%%%%%%%%%%%%%%%%%%%%%%%%%%%
\subsection{States bifurcating from the ground state}
%%%%%%%%%%%%%%%%%%%%%%%%%%%%%%%%%%%%%%%%%%%%%%%%%%%%%%%%%%%%%%%%%%%%%%%%%%%

We also explored states bifurcating from the ground state. In particular,
at $\mu\approx 0.560$ a double-ring solution bifurcates away from the ground
state. This double-ring (see panels (a) in Fig.~\ref{Ring_Exci}) contains
the ground state populating the central well coupled to two {\em out-of-phase} 
rings, that populate the ring portion of the external potential, as can be 
seen in the top panel of Fig.~\ref{Ring_Exci}(a) ---depicting the real part 
of the solution--- where the phase difference between the inner and outer 
rings is evident. 
Namely, this state effectively contains
a ring dark soliton~\cite{gtheo,2010JPhA.43u3001F,doi:10.1137/1.9781611973945}
inside the outside ring channel.
Figure~\ref{Gr_Spectra} depicts the BdG spectra for the ground state and
the double-ring state as a function of $\mu$. As expected, the ground state
is always (neutrally) stable. However, as it is clear from the figure,
the double ring is unstable since its inception. It is relevant to note
that this has been recently demonstrated to be generically the case
due to their azimuthal undulations in the presence of an external
radial potential with the quadrupolar undulations representing the
first among such spatial modes that becomes unstable~\cite{PhysRevLett.118.244101}.

Interestingly, there exist further states bifurcating in turn from the double-ring 
solution. These states, depicted in panels (b)--(d)
in Fig.~\ref{Ring_Exci}, correspond to the double-ring with out-of-phase ``petals'' 
along the azimuthal direction.
The bifurcation progression of the double-ring from the ground state and, 
subsequently, the states bifurcating from the double-ring is more evidently 
portrayed in Fig.~\ref{Number_atoms_Ring} that depicts the particle numbers
for these solutions as a function of $\mu$.
It is relevant to note that, apparently, configurations with higher number
of petals bifurcate first from the double-ring.
This bifurcation cascade continues beyond what is shown in Fig.~\ref{Ring_Exci}
(where only the first couple of bifurcating branches are depicted).

%%%%%%%%%%%%%%%%%%%%%%%%%%%%%%%%%%%%%%%%%%%%%%%%%%%%%%%%%%%%%%%%%%%%%%%%%%%
\begin{figure}[!htb]
\includegraphics[width=\columnwidth]{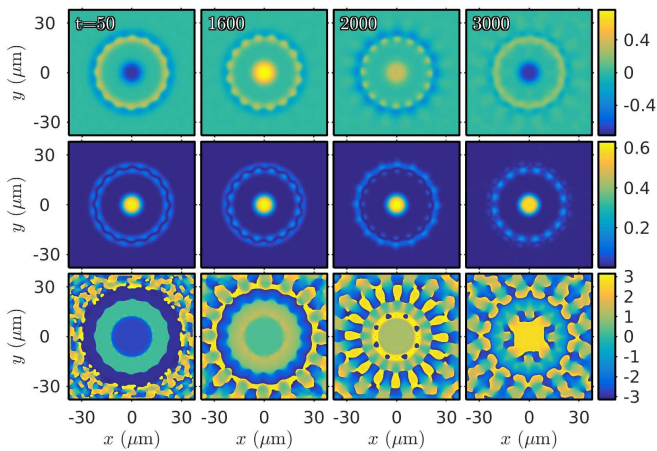}
\caption{(Color online)
Evolution of the double-ring configuration (see panels (a) in Fig.~\ref{Ring_Exci}) 
heavily perturbed (30 times the normalized eigenvector) 
with an eigenvector picked from the third instability in the 
BdG spectra (see bottom panels in Fig.~\ref{Gr_Spectra}). 
More precisely, we perturb the double-ring state solution 
with the eigenvector of the ring state calculated for $\mu=0.63$.  
The top, middle, and bottom rows of panels display, respectively, 
the real part, the density, and phase of the profiles at the times indicated.
In this figure, as is the case in all the figures in this work, the
indicated times are measured in non-dimensional units as per
the adimensionalization discussed below Eq.~(\ref{NLS_eq}).
}
\label{Sol_heavy_pert_ring}
\end{figure}
%%%%%%%%%%%%%%%%%%%%%%%%%%%%%%%%%%%%%%%%%%%%%%%%%%%%%%%%%%%%%%%%%%%%%%%%%%%

As concerns the stability of the bifurcating states, it is important
to stress that the double-ring solution is unstable since its
emergence from the ground state around $\mu\simeq0.560$ and, therefore,
all the subsequent bifurcating states from the double-ring inherit the instability
from their double ring ``ancestor'' and are thus always unstable as well.
Furthermore, it is interesting to note that the the first few instabilities seen 
in the BdG spectrum of the double ring (bottom-right panel in Fig.~\ref{Gr_Spectra}) 
 coincide with the critical mass values corresponding to the emergence
 of the different bifurcating states from the double-ring configuration.
 Another way to state this in the language of dynamical systems is
 that these multi-petal states are emerging via supercritical
 pitchfork (symmetry breaking) bifurcations, leading to the
 further destabilization of the radially
 symmetric state via the emergence of
 a wide variety of azimuthally modulated ones.

Finally, in order to monitor the evolution of instabilities for the
double-ring, we depict in Fig.~\ref{Sol_heavy_pert_ring} the dynamical 
destabilization of the double-ring.
In this case, we perturb the double-ring profile with an eigenvector picked from the 
third instability in the BdG spectra (see bottom-right panel in Fig.~\ref{Gr_Spectra}). The wave form involving the relevant wavenumber is
clearly dynamically amplified and eventually destroys the ring
like structure in favor of one that bears the periodicity of
the imposed perturbation.
%

%%%%%%%%%%%%%%%%%%%%%%%%%%%%%%%%%%%%%%%%%%%%%%%%%%%%%%%%%%%%%%%%%%%%%%%%%%%
\begin{figure}[!htb]
\includegraphics[width=\columnwidth]{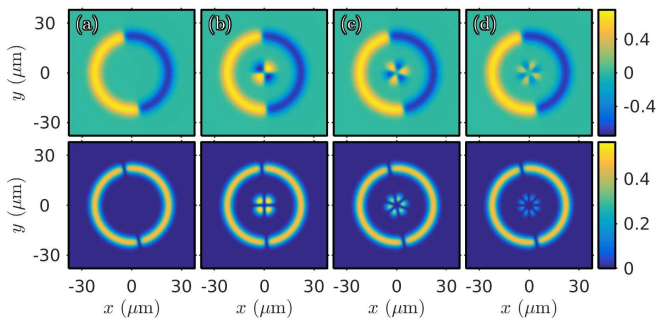}
\caption{(Color online)
2-dark soliton profile and its bifurcating states for $\mu=0.7$.
Same layout as in previous figures.
The corresponding particle numbers as a function of $\mu$ are 
depicted in Fig.~\ref{Number_atoms_2SoliFunda}.
}
\label{2Soli_Funda}
\end{figure}
%%%%%%%%%%%%%%%%%%%%%%%%%%%%%%%%%%%%%%%%%%%%%%%%%%%%%%%%%%%%%%%%%%%%%%%%%%%

%%%%%%%%%%%%%%%%%%%%%%%%%%%%%%%%%%%%%%%%%%%%%%%%%%%%%%%%%%%%%%%%%%%%%%%%%%%
\begin{figure}
\includegraphics[width=0.85\columnwidth]{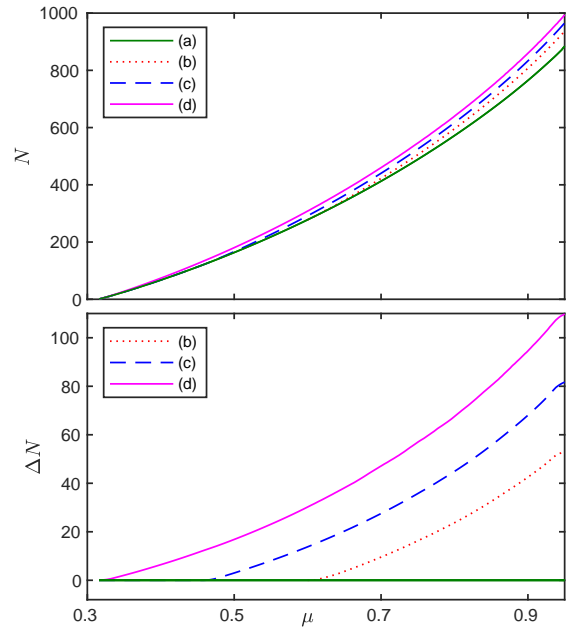}
\caption{(Color online)
Top: Particle number $N$ as a function of $\mu$ for the stationary states 
bifurcating from the 2-soliton solution (a). 
Bottom: Particle number difference $\Delta N$ between the states
bifurcating from the 2-soliton configuration and the 2-soliton
configuration itself.
The corresponding profiles are depicted in Fig.~\ref{2Soli_Funda}.
The first three bifurcating states from the 2-soliton solution (a)
bifurcate at:
(b) $\mu\simeq 0.321$,
(c) $\mu\simeq 0.466$, and
(d) $\mu\simeq 0.614$.
}
\label{Number_atoms_2SoliFunda}
\end{figure}
%%%%%%%%%%%%%%%%%%%%%%%%%%%%%%%%%%%%%%%%%%%%%%%%%%%%%%%%%%%%%%%%%%%%%%%%%%%

%%%%%%%%%%%%%%%%%%%%%%%%%%%%%%%%%%%%%%%%%%%%%%%%%%%%%%%%%%%%%%%%%%%%%%%%%%%
\begin{figure}[!htb]
\includegraphics[width=0.85\columnwidth]{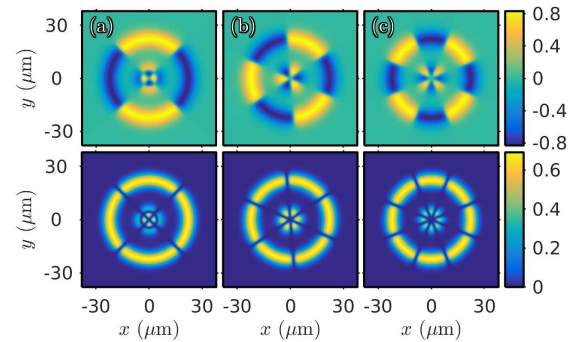}
\caption{(Color online)
Same as Fig.~\ref{2Soli_Funda}, but showing (from left to right) the first 
state bifurcating from the 4-, 6-, and 8-soliton profiles.}
\label{Con_468}
\end{figure}
%%%%%%%%%%%%%%%%%%%%%%%%%%%%%%%%%%%%%%%%%%%%%%%%%%%%%%%%%%%%%%%%%%%%%%%%%%%

%%%%%%%%%%%%%%%%%%%%%%%%%%%%%%%%%%%%%%%%%%%%%%%%%%%%%%%%%%%%%%%%%%%%%%%%%%%
\subsection{States bifurcating from the $n$-dark soliton configurations}
%%%%%%%%%%%%%%%%%%%%%%%%%%%%%%%%%%%%%%%%%%%%%%%%%%%%%%%%%%%%%%%%%%%%%%%%%%%

In a similar manner as we identified bifurcating states from the
ground state and subsequently from the double-ring
in the previous section, we now follow the bifurcating states from the
$n$-dark soliton solutions and their associated phenomenology.
For instance, Fig.~\ref{2Soli_Funda} depicts, alongside the 2-soliton
solution, its first three bifurcating states.
In this case the bifurcating states pertain to excitations of the central
well of the external potential. These central well excitations correspond to
azimuthal, out-of-phase, ``multi-petal'' configurations.
Figure~\ref{Number_atoms_2SoliFunda} depicts the particle numbers for these
configurations. In particular, the bottom panel displays the particle number difference $\Delta N$ between the central excited states and the original 2-soliton solution. When this diagnostic departs from zero,
it signals the emergence of a bifurcation of a new branch from a
previously existing one (with $\Delta N=0$).
As shown in Fig.~\ref{Con_468}, similarly to the bifurcating states from the 2-soliton configuration, we were able to identify bifurcating states from the 4-, 6-, and 8-soliton configurations.

%%%%%%%%%%%%%%%%%%%%%%%%%%%%%%%%%%%%%%%%%%%%%%%%%%%%%%%%%%%%%%%%%%%%%%%%%%%
\begin{figure}[!htb]
\includegraphics[width=1.00\columnwidth]{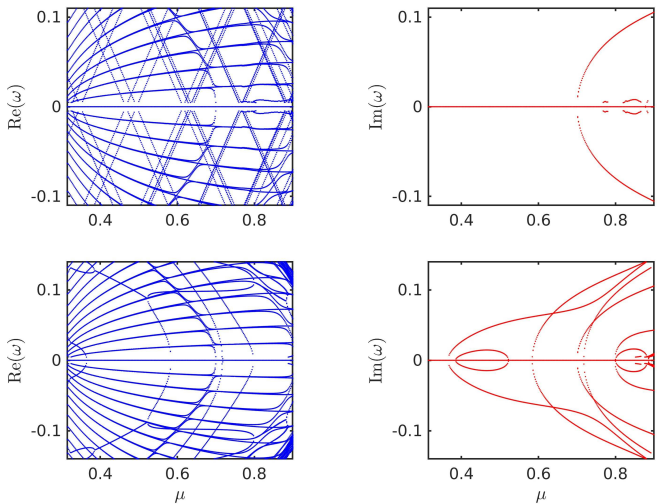}
\caption{(Color online)
Stability BdG spectra for the 2-soliton configuration and its first 
bifurcating state as a function of $\mu$.
Same layout as in Fig.~\ref{Gr_Spectra}.
The corresponding profiles for $\mu=0.7$ are depicted, respectively, in the 
first two columns of Fig.~\ref{2Soli_Funda}.
}
\label{2Sol_Spectra}
\end{figure}
%%%%%%%%%%%%%%%%%%%%%%%%%%%%%%%%%%%%%%%%%%%%%%%%%%%%%%%%%%%%%%%%%%%%%%%%%%%

Now that we have identified the 2-soliton solution and its bifurcating states, 
let us briefly discuss the ensuing stability as a function of $\mu$. 
In Fig.~\ref{2Sol_Spectra} we depict the BdG spectra of the 2-soliton state 
(top row) together with the spectra of its first bifurcating states (bottom row)
---profiles for these configurations for $\mu=0.7$ are depicted in the first two columns 
of Fig.~\ref{2Soli_Funda}.
The BdG spectrum for the 2-soliton configuration indicates that this profile
is (neutrally) stable for $\mu < 0.702$. 
For larger values of the chemical potential (not shown here) other instabilities 
arise, however we do not consider them here given their much weaker growth rates. 

\begin{figure*}[!htb]
\includegraphics[width=0.95\textwidth]{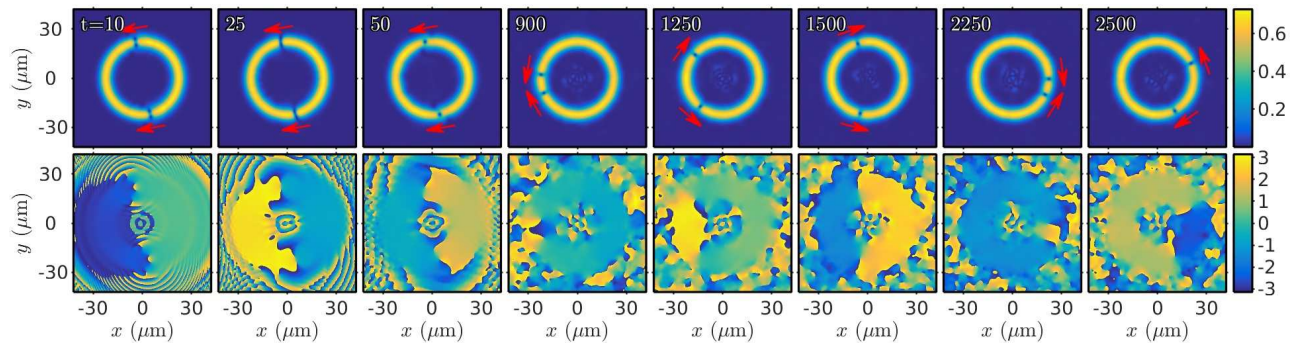}
\caption{(Color online)
Density (top row of panels) and phase (bottom row of panels) plots showing the time 
evolution of the 2-soliton ground state heavily perturbed (30 times the normalized eigenvector)
with an eigenvector picked 
from the second instability in the BdG spectra (see Fig.~\ref{2Sol_Spectra}). 
Specifically, we perturb a 2-soliton configuration obtained for $\mu=0.904$ with 
the second eigenvector of the 2-soliton state calculated for $\mu=0.93$. 
We confirmed the same type of dynamics when adding smaller perturbations. 
}
\label{2Sol_heavy_pert_br2}
\end{figure*}

For instance, Fig.~\ref{2Sol_heavy_pert_br2} shows the long time evolution of 
the 2-soliton ground state heavily perturbed with an eigenvector picked from the
second instability in the BdG spectrum. 
We observe that, when perturbed, the 2-soliton configuration develops two pairs of 
vortices which travel inside the ring. The vortex nature of these traveling localized
solutions becomes apparent in the phase plots (see bottom row of panels)
and the corresponding 2$\pi$ winding at the vortex locations.
To guide the reader we have included (red) arrows that indicate the direction of
motion for the vortices.
As time progresses one of the vortices in each vortex pair gets ``absorbed'' by the
edge of the ring ($t\approx 50$) leaving only two vortices of opposite charge to
run along the ring. The two vortices travel towards each other, then reverse direction, 
move again towards each other, bounce off again etc. The vortices are found to move 
back and forth for a prolonged time before they annihilate for longer times (not
shown here) and the configuration settles down to a slightly perturbed ring
without any apparent vortices in the bulk of the ring. 

Here, we omit the time evolution of instabilities corresponding to the higher excited
states of the 2-soliton configuration
since they do not provide any new insights into the dynamical properties. 
In all cases, vortices are found to travel back and forth inside the ring. 
For the excited states of the 2-soliton configuration, we also observe that vortices are created
in the central portion of the cloud. However, those might be less relevant for experiments
as the density is low there and the vortices are more tightly packed.

%%%%%%%%%%%%%%%%%%%%%%%%%%%%%%%%%%%%%%%%%%%%%%%%%%%%%%%%%%%%%%%%%%%%%%%%%%%
%%%%%%%%%%%%%%%%%%%%%%%%%%%%%%%%%%%%%%%%%%%%%%%%%%%%%%%%%%%%%%%%%%%%%%%%%%%

%%%%%%%%%%%%%%%%%%%%%%%%%%%%%%%%%%%%%%%%%%%%%%%%%%%%%%%%%%%%%%%%%%%%%%%%%%%
\begin{figure}[t]
\includegraphics[width=1.00\columnwidth]{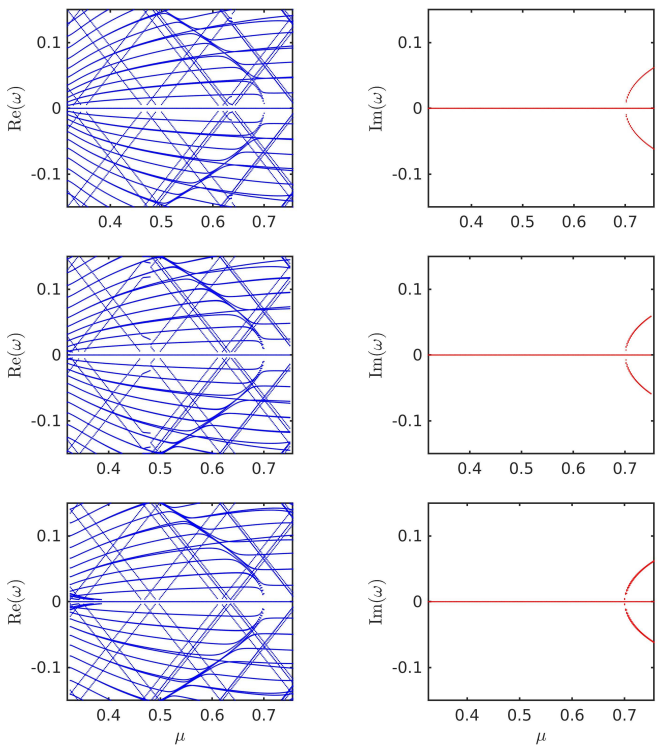}
\caption{(Color online)
Stability BdG spectra for the 4-, 6- and 8-soliton states (from top to bottom).
Same layout as in Fig.~\ref{Gr_Spectra}. 
The corresponding profiles are depicted in the panels (c)--(e) of
Fig.~\ref{Soli_Funda}.
}
\label{MultiSol_BdG_Spectra}
\end{figure}
%%%%%%%%%%%%%%%%%%%%%%%%%%%%%%%%%%%%%%%%%%%%%%%%%%%%%%%%%%%%%%%%%%%%%%%%%%%

%%%%%%%%%%%%%%%%%%%%%%%%%%%%%%%%%%%%%%%%%%%%%%%%%%%%%%%%%%%%%%%%%%%%%%%%%%%
\begin{figure*}[!htb]
\includegraphics[width=0.85\textwidth]{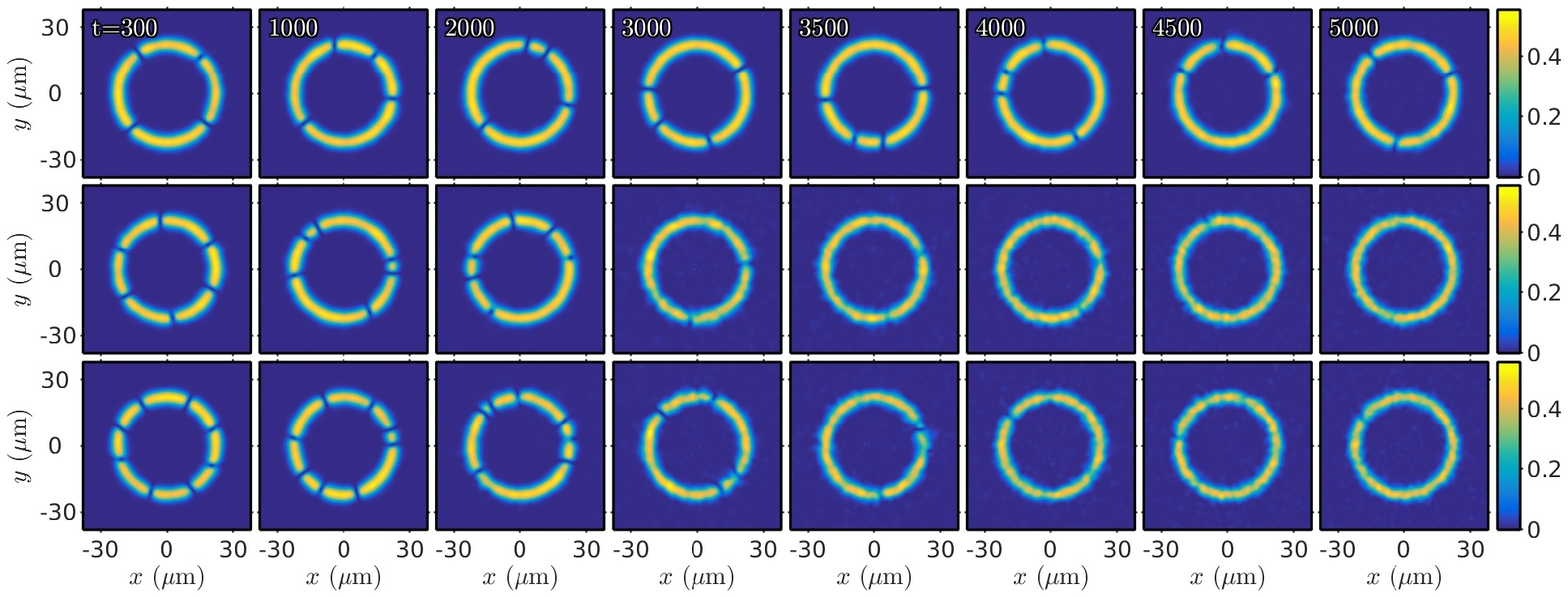}
\caption{(Color online)
Evolution dynamics for the 4-soliton (top row), 6-soliton (middle row), and 
8-soliton (bottom row) configurations heavily perturbed (30 times the normalized eigenvector)
with an eigenvector picked 
from the first instability of the corresponding BdG spectra.
In all cases we perturb the $n$-soliton state obtained for $\mu=0.7$ 
with the eigenvector of the $n$-soliton state calculated for $\mu=0.71$.
}
\label{468Sol_heavy_pert}
\end{figure*}
%%%%%%%%%%%%%%%%%%%%%%%%%%%%%%%%%%%%%%%%%%%%%%%%%%%%%%%%%%%%%%%%%%%%%%%%%%%

For completeness, we depict in Fig.~\ref{MultiSol_BdG_Spectra} the BdG stability spectra 
for the 4-, 6- and 8-soliton solutions. 
As it was the case for the 2-soliton configuration, the $n$-soliton configurations
are also stable for $\mu < 0.702$ and the spectra are quite similar. This is
straightforward to understand as the corresponding dark solitons are placed relatively
far away from each other along the ring and, therefore, their mutual interaction
is (exponentially) weak and thus not very noticeable when dealing with a handful
of solitons. Nonetheless, higher-order excited states including a large number of
dark solitons will correspond to relatively shorter mutual separations leading
to stronger interactions and modifications of the stability spectra. We defer
the study of such cases to future publications.

Finally, we depict in Fig.~\ref{468Sol_heavy_pert} the corresponding dynamical
evolution for the $n$-soliton profiles for $n=4$, 6, and 8, when perturbed 
with eigenvectors picked from the first instability in their BdG spectra.
Note that in all cases the dynamics tends to
lead to the disintegration of the dark solitons (through collisions and/or splitting 
into vortex pairs that in turn get ``absorbed'' by the periphery of the ring). 
Eventually, and potentially after long transient
stages, the evolution settles into a perturbed ring structure without dark solitons or vortices in its bulk. 
%

%%%%%%%%%%%%%%%%%%%%%%%%%%%%%%%%%%%%%%%%%%%%%%%%%%%%%%%%%%%%%%%%%%%%%%%%%%%
\begin{figure*}[!htb]
\includegraphics[width=0.85\textwidth]{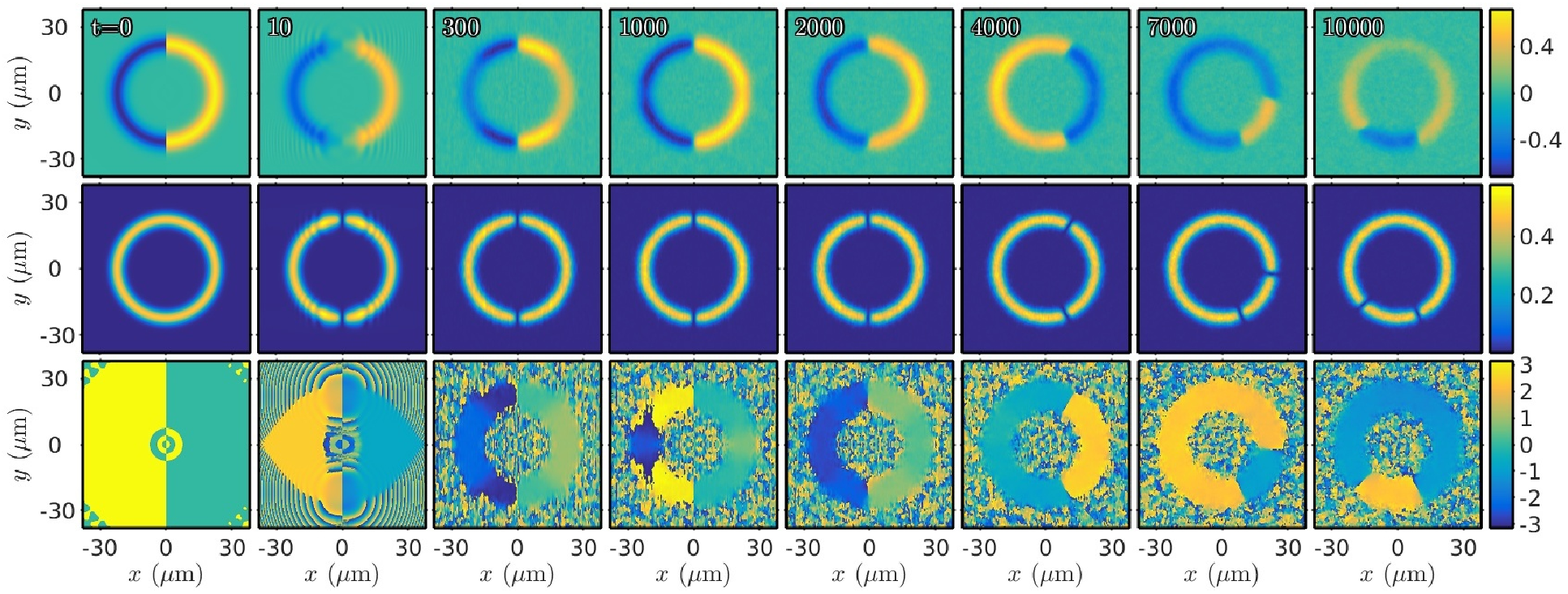}
\caption{(Color online)
Dynamics ensuing from the phase imprinting the 2-soliton configuration
for $\mu=0.9$.
}
\label{fig:imprint2sol}
\end{figure*}
%%%%%%%%%%%%%%%%%%%%%%%%%%%%%%%%%%%%%%%%%%%%%%%%%%%%%%%%%%%%%%%%%%%%%%%%%%%

%%%%%%%%%%%%%%%%%%%%%%%%%%%%%%%%%%%%%%%%%%%%%%%%%%%%%%%%%%%%%%%%%%%%%%%%%%%
\begin{figure*}[!htb]
\includegraphics[width=0.85\textwidth]{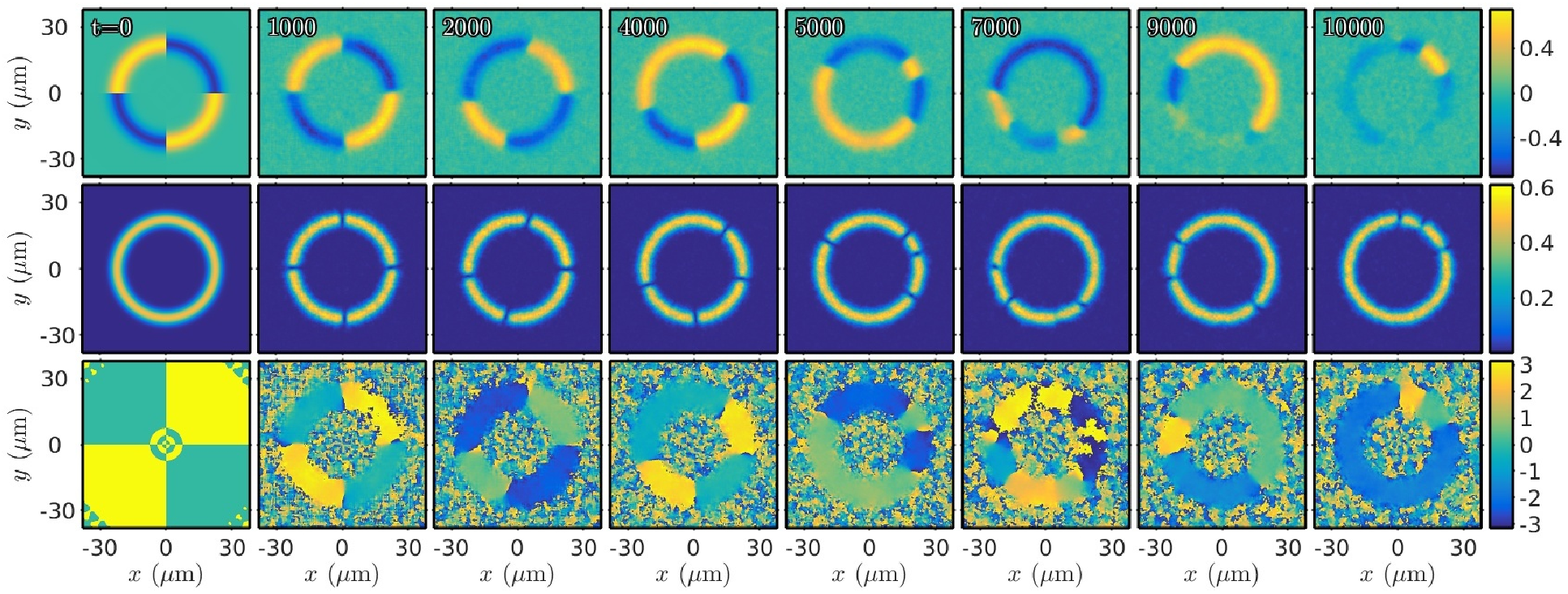}
\caption{(Color online)
Same as in Fig.~\ref{fig:imprint2sol} but for the 4-soliton configuration.
}
\label{fig:imprint4sol}
\end{figure*}
%%%%%%%%%%%%%%%%%%%%%%%%%%%%%%%%%%%%%%%%%%%%%%%%%%%%%%%%%%%%%%%%%%%%%%%%%%%

%%%%%%%%%%%%%%%%%%%%%%%%%%%%%%%%%%%%%%%%%%%%%%%%%%%%%%%%%%%%%%%%%%%%%%%%%%%
\subsection{Phase imprinting of $n$-dark soliton states}
%%%%%%%%%%%%%%%%%%%%%%%%%%%%%%%%%%%%%%%%%%%%%%%%%%%%%%%%%%%%%%%%%%%%%%%%%%%

We now explore the especially important ---in terms of
a practical implementation--- possibility of seeding in the experiment 
some of the excited state configurations that we described above.
In particular, we are interested in the experimental possibility of
initializing configurations that bear $n$-dark solitons and let them 
evolve to study their interactions and collisional dynamics.
For that purpose, we start with the ring steady state depicted in 
Fig.~\ref{Soli_Funda}(b). As mentioned above, this solution
exists for $\mu\geq 0.313$ and it is stable for $\mu < 0.702$ and therefore
it is a good candidate to be attainable in a physical experiment.
Then, by using a phase imprinting technique, e.g., by shining 
laser light on one half of the condensate for a short period of 
time~\cite{Denschlag97,PhysRevLett.83.5198,PhysRevA.65.043611,PhysRevA.66.033602}, 
whereby half of the ring's phase is shifted by $\pi$ with respect to the other 
half, it is possible to generate an initial condition that has the correct phase
profile of a 2-dark soliton state. Such scenarios with multiple
phase jumps have been previously used in quasi-1d settings
in order to examine the effectively 1d interaction of dark solitary
waves~\cite{beckerr}.

We have tested that this technique is successful at seeding $n$-dark 
solution solutions for chemical potentials below the instability threshold
around $\mu\approx0 .7$ (results not shown here).
However, as we are interested not only in seeding steady states
in the experiments, but also in observing the potentially unstable dynamics 
of these $n$-dark soliton solutions. In that light,
we focus our attention here
on phase imprinting $n$-dark soliton solutions past their stability
threshold (i.e., $\mu$ a bit larger than 0.7).
This is precisely what  is depicted in Fig.~\ref{fig:imprint2sol}
where the initial condition (first column of panels) corresponds the ring
steady state with a phase imprinting such that the phase of the
left half is $+\pi$ while the phase of the right half is $0$.
As  can be observed from the figure, after an initial period of adjustment
($t<300$), where the imprinted phase forces the dark soliton nucleation, a pair 
of dark solitons on opposite sides of the ring is formed. This configuration
corresponds to a slightly perturbed 2-dark soliton state. This state, being
unstable for the chosen value of $\mu$ as per the discussion in the previous 
sections, evolves in a manner akin to the one depicted in 
Fig.~\ref{2Sol_heavy_pert_br2}. Namely, the dark solitons start moving and 
colliding along the ring.

This phase-imprinting technique can be straightforwardly generalized to
higher number of dark solitons by imprinting the appropriate phase. For instance,
by imprinting a phase difference across the horizontal axis and then doing
the same across the vertical axis, one is left with the appropriate phase
to nucleate the $4$-dark soliton state. This case is depicted in 
Fig.~\ref{fig:imprint4sol} whose dynamical evolution in now similar to
the one depicted in the first row of panels in Fig.~\ref{468Sol_heavy_pert}.
It is relevant to mention that the dynamics of the unstable $n$-dark soliton
eventually leads to a perturbed ground state as the dark solitons 
destabilize towards the formation of vortex pairs, which in turn scatter
and ultimately get absorbed by the periphery
of the ring.
It is natural to expect
that as the ring gets thinner and more quasi-one-dimensional
the relevant states will be progressively stabilized against
such transverse undulations and the associated breakup towards
vortex dipoles~\cite{brandas}.

%%%%%%%%%%%%%%%%%%%%%%%%%%%%%%%%%%%%%%%%%%%%%%%%%%%%%%%%%%%%%%%%%%%%%%%%%%%
\begin{figure*}[!htb]
\includegraphics[width=0.90\textwidth]{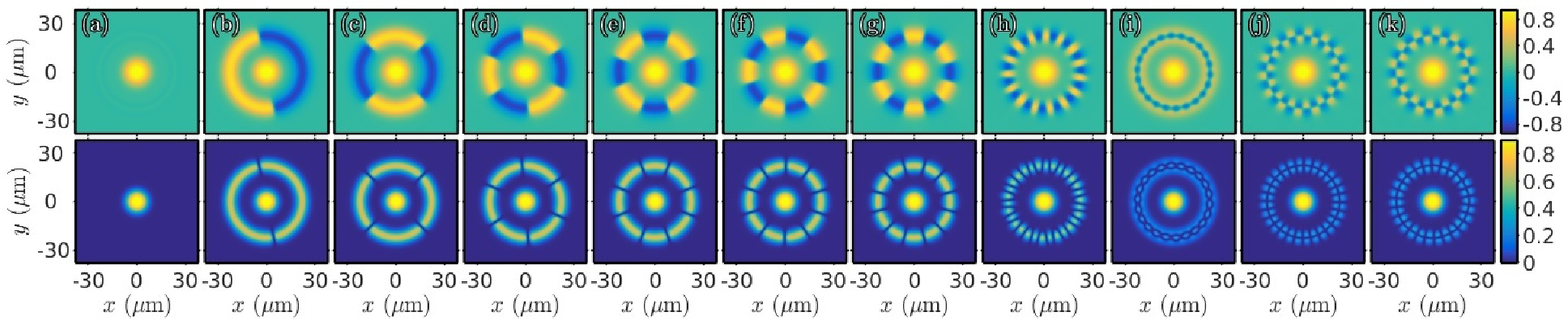}
\caption{(Color online)
Real states bifurcating from the ground state for $\mu=0.9$
[except $\mu=0.96$ for panel (k)].
}
\label{Ground_real_states}
\end{figure*}
%%%%%%%%%%%%%%%%%%%%%%%%%%%%%%%%%%%%%%%%%%%%%%%%%%%%%%%%%%%%%%%%%%%%%%%%%%%

%%%%%%%%%%%%%%%%%%%%%%%%%%%%%%%%%%%%%%%%%%%%%%%%%%%%%%%%%%%%%%%%%%%%%%%%%%%
\begin{figure*}[!htb]
\includegraphics[width=0.90\textwidth]{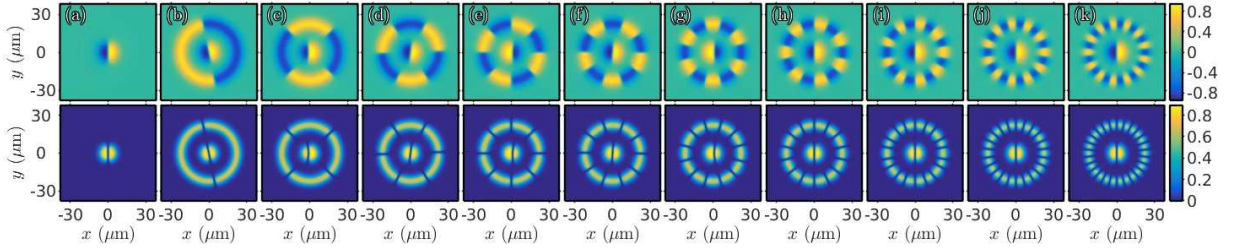}
\caption{(Color online)
Real states calculated from the dipole state for $\mu=0.9$
}
\label{Dipole_real_states}
\end{figure*}
%%%%%%%%%%%%%%%%%%%%%%%%%%%%%%%%%%%%%%%%%%%%%%%%%%%%%%%%%%%%%%%%%%%%%%%%%%%

%%%%%%%%%%%%%%%%%%%%%%%%%%%%%%%%%%%%%%%%%%%%%%%%%%%%%%%%%%%%%%%%%%%%%%%%%%%
\begin{figure*}[!htb]
\includegraphics[height=4.3cm]{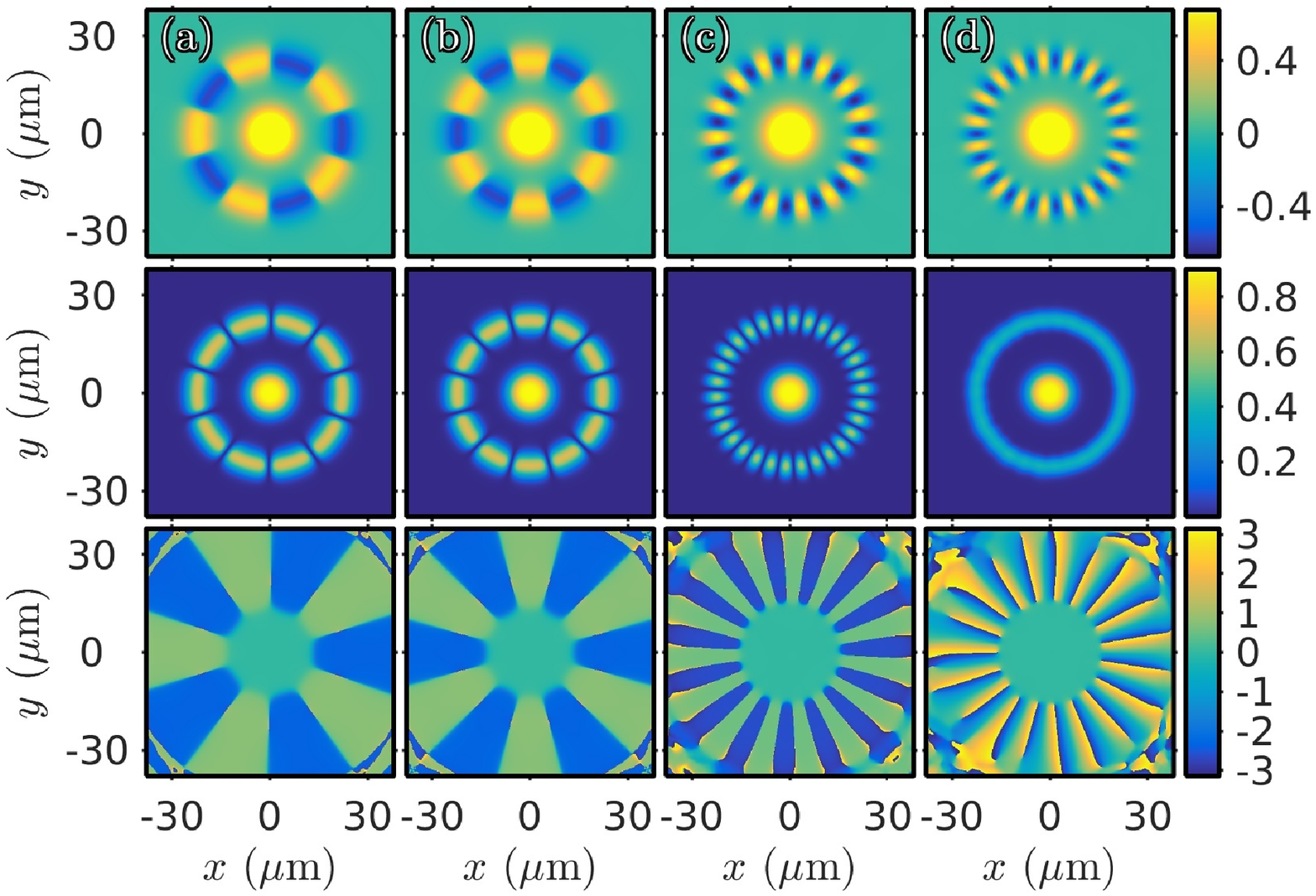}
\includegraphics[height=4.3cm]{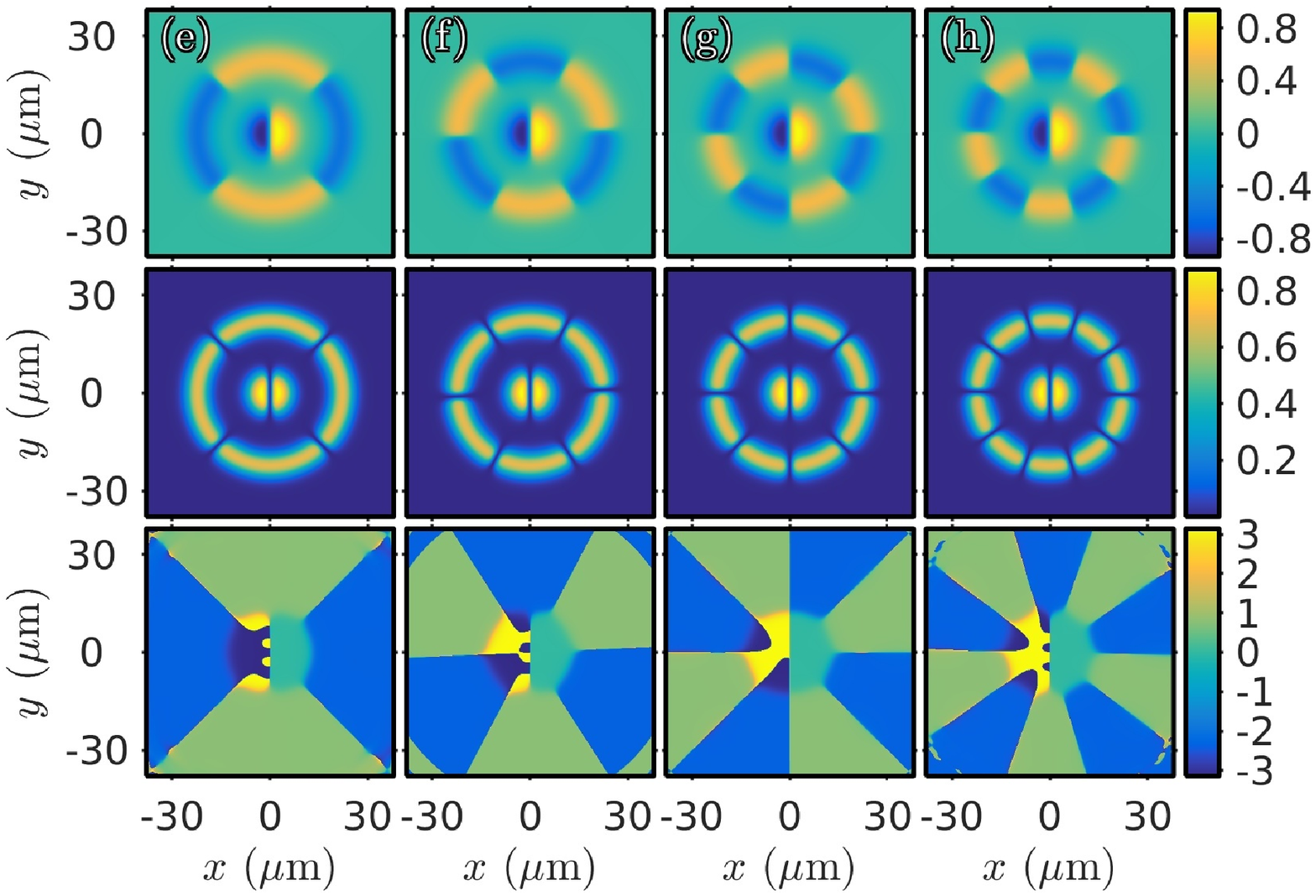}
\includegraphics[height=4.3cm]{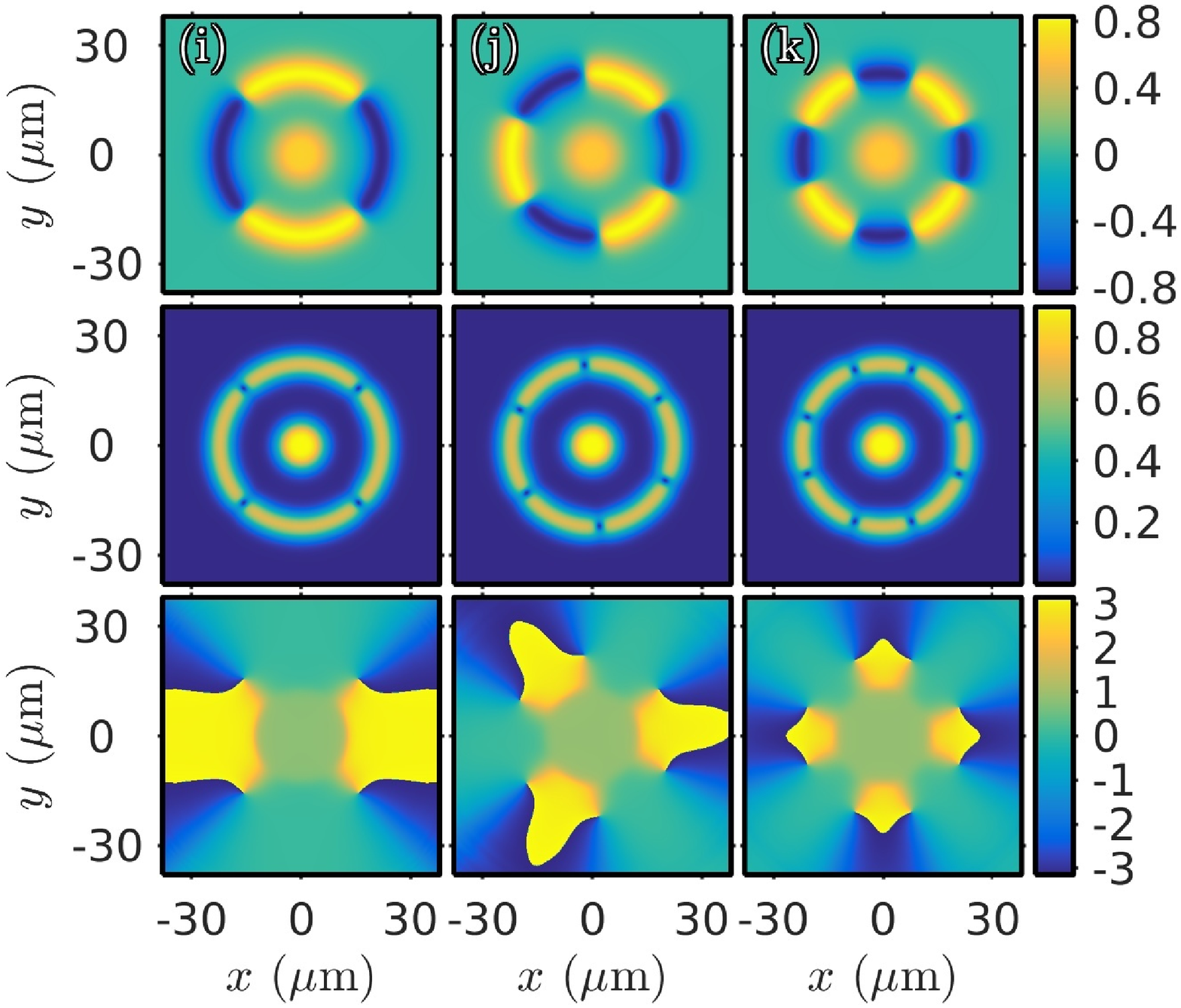}
\caption{(Color online)
(a)--(d) Complex states calculated from the ground state.
These profiles correspond to an $n$-dark soliton state coupled
to a the ground state.
(e)--(h) Complex states calculated from the dipole state.
These profiles correspond to an $n$-dark soliton state coupled
to a dipole state at the center of the cloud.
(i)--(k) Vortex like states calculated from the ground state.
These states a similar to the ones depicted in panels (a)--(d)
by replacing the $n$-dark soliton state by a ring of $n$ vortices.
$\mu=0.9$ in all cases.
}
\label{Complex_states}
\end{figure*}
%%%%%%%%%%%%%%%%%%%%%%%%%%%%%%%%%%%%%%%%%%%%%%%%%%%%%%%%%%%%%%%%%%%%%%%%%%%

%%%%%%%%%%%%%%%%%%%%%%%%%%%%%%%%%%%%%%%%%%%%%%%%%%%%%%%%%%%%%%%%%%%%%%%%%%%
\subsection{A Zoo of More Exotic States }
%%%%%%%%%%%%%%%%%%%%%%%%%%%%%%%%%%%%%%%%%%%%%%%%%%%%%%%%%%%%%%%%%%%%%%%%%%%

In addition to the states we constructed from the linear limit, there also
exist 
states which bifurcate from the ground and dipole states and their excitations.
Appropriate initial guesses for these states have been constructed by using 
the well-known ground and dipole ans\"atze for solutions of Eq.~(\ref{steady}) 
in the presence of a harmonic external potential. 
For instance, as depicted in Fig.~\ref{Ground_real_states}, 
there is a plethora of states bifurcating from the ground state. All of the states presented in this figure are real and pertain the combination of an $n$-dark soliton
solution (populating the ring) coupled to a phase-less hump of mass localized
in the central well (namely, the remnant of the ground state of the system).
We have checked that all of these states are actually unstable (results
not shown here).
Similarly, as depicted in Fig.~\ref{Dipole_real_states},
it is possible to find more families of purely real solutions 
corresponding to the combination of, again, an $n$-dark soliton solution 
(populating the ring) but now coupled to the first excited state of the 
ground state (namely, the dipole consisting of a plus-minus hump at the 
center of the cloud).  We have also checked that all of these states are 
actually unstable (results not shown here).
This process can be extended for higher excited states of the ground state
coupled to the $n$-dark soliton configuration on the ring.

Furthermore, it is also possible to find rich families of genuinely complex
solutions. For instance, as seen in panels (a)--(d) of Fig.~\ref{Complex_states},
it is possible to couple the $n$-dark soliton state with the ground state
with a non-trivial phase difference between these two states.
In the same vein, as is shown in  panels (e)--(h) in Fig.~\ref{Complex_states},
it is possible to couple with a non-trivial relative phase the $n$-dark 
soliton state with the dipole state at the center of the cloud.
We have also checked that all of these states are always unstable 
(results not shown here).

Finally, it is relevant to mention that non-trivial phase configurations
can be constructed by replacing the $n$-dark soliton solutions on the
ring by a necklace of $n$-vortex solutions.
These more exotic profiles are depicted in panels (i)--(k) in 
Fig.~\ref{Complex_states} for the case of 2, 4, and 8 vortices,
respectively.
%

%%%%%%%%%%%%%%%%%%%%%%%%%%%%%%%%%%%%%%%%%%%%%%%%%%%%%%%%%%%%%%%%%%%%%%%%%%%
\section{Conclusions and Future Challenges}
\label{Con}
%%%%%%%%%%%%%%%%%%%%%%%%%%%%%%%%%%%%%%%%%%%%%%%%%%%%%%%%%%%%%%%%%%%%%%%%%%%

We have studied the stationary and dynamical properties of BEC profiles
supported by a ring-shaped potential with a target-like profile that
has been used in a number of recent experiments conducted at NIST~\cite{eckel_2013,PhysRevA.92.033602}. 
By following steady states and their bifurcations from the linear
(low atom number) limit, we 
have obtained a wide range of solution branches (not all of which were
shown here) and studied the corresponding 
stability properties as the chemical potential $\mu$ (cf.~atom number) 
is varied. Importantly, numerous among these states were found
to be potentially stable, including states carrying multiple (2-,
4-, 6-, 8-) solitons between the starting point of the respective
branches and up to a suitable critical value of the chemical potential.
Past this critical $\mu$ value, we studied the ensuing dynamics of the
dark solitons around the ring. We typically observed that the dark solitons
bounce back-and-forth in the ring until they disappear in a process
involving each dark soliton splitting into a vortex pair and then the
vortices getting eventually absorbed by the periphery of the ring.
This process eventually led to a weakly perturbed (i.e, almost homogeneous) 
ring void of any dark solitons or vortices that persisted for long times.

In the case of $n$-dark soliton solutions, taking advantage of
their spectral stability, we illustrated their potential
for experimental realization by using 
phase-imprinting techniques to seed them in the condensate.
We were not only able to seed stable $n$-dark soliton solutions but, equally
interestingly, to seed unstable solutions whereby the ensuing dark soliton
instability dynamics can be studied.

Additionally, a plethora of states was identified involving
a combination of (ground or excited) states supported 
by the central well of the target-like potential coupled with states
supported by the ring channel. 
The states supported by the central well corresponded to the trivial-phase
ground state and its excitations in the form of dipole, quadrupole, 
etc.~states.
On the other hand, the ring channel accepts $n$-dark (equidistant) soliton 
solutions where $n$ is even as the periodicity of the ring enforces an even
number of dark solitons.
We also followed states that, instead of bifurcating from the linear
limit, bifurcate from the ground state of the system (a phase-less
hump populating the central well). These states correspond to
double-ring, out-of-phase, solutions and ``petal''-like patterns
around the ring. 

It would be interesting to implement the phase-imprinting methodology
in the actual experiment as it would naturally allow for the study
of dark soliton dynamics and interactions especially so in such
an annular setup. The potential control of the spatial width of the
annulus and the associated control of the snaking stability of the solitonic
structures could play a significant role in the explored dynamics.
From the modeling perspective it would be interesting to study
the stability and dynamics of steady states bearing a large number
of dark solitons. For instance, it is known that a chain of dark
solitons can be approximated by a Toda lattice on the solitons'
positions and thus one can create (Toda) solitons riding on a backbone 
of dark solitons (see Ref.~\cite{Manjun} and references therein).
Furthermore, a systematic extension of the present studies considering
the vortex patterns in the present setting would naturally complement
the present solitonic considerations.
Lastly, considering extensions of this type of set up
also in higher dimensions and suitable (e.g. toroidal-poloidal)
geometries may be particularly interesting and relevant in its
own right, as well as an appreciation of which (potentially vortical)
patterns may be dynamically stable. 
%

%%%%%%%%%%%%%%%%%%%%%%%%%%%%%%%%%%%%%%%%%
\section*{Acknowledgements}
%%%%%%%%%%%%%%%%%%%%%%%%%%%%%%%%%%%%%%%%%

Some of the work of M.H.~was undertaken as a visiting research scholar at the 
Department of Mathematics and Statistics, University of Massachusetts, employed 
by the University of Oldenburg and financially supported by FP7, Marie Curie 
Actions, People, International Research Staff Exchange Scheme (IRSES-605096).
P.G.K.\ and R.C.G.\ and M.A.E.\ gratefully acknowledge the support from the 
National Science Foundation, under grants PHY-1602994, PHY-1603058, and PHY-1707776.


\begin{thebibliography}{XX}

\expandafter\ifx\csname natexlab\endcsname\relax\def\natexlab#1{#1}\fi
\expandafter\ifx\csname bibnamefont\endcsname\relax
  \def\bibnamefont#1{#1}\fi
\expandafter\ifx\csname bibfnamefont\endcsname\relax
  \def\bibfnamefont#1{#1}\fi
\expandafter\ifx\csname citenamefont\endcsname\relax
  \def\citenamefont#1{#1}\fi
\expandafter\ifx\csname url\endcsname\relax
  \def\url#1{\texttt{#1}}\fi
\expandafter\ifx\csname urlprefix\endcsname\relax\def\urlprefix{URL }\fi
\providecommand{\bibinfo}[2]{#2}
\providecommand{\eprint}[2][]{\url{#2}}

\bibitem[{\citenamefont{Pethick and Smith}(2008)}]{pethick_smith_2008}
\bibinfo{author}{\bibfnamefont{C.~J.} \bibnamefont{Pethick}} \bibnamefont{and}
  \bibinfo{author}{\bibfnamefont{H.}~\bibnamefont{Smith}},
  \emph{\bibinfo{title}{Bose-Einstein Condensation in Dilute Gases}}
  (\bibinfo{publisher}{Cambridge University Press}, \bibinfo{year}{2008}),
  \bibinfo{edition}{2nd} ed.

\bibitem[{\citenamefont{Pitaevskii and Stringari}(2003)}]{BEC_Stringari}
\bibinfo{author}{\bibfnamefont{L.}~\bibnamefont{Pitaevskii}} \bibnamefont{and}
  \bibinfo{author}{\bibfnamefont{S.}~\bibnamefont{Stringari}},
  \emph{\bibinfo{title}{Bose-Einstein Condensation}}
  (\bibinfo{publisher}{Oxford University Press}, \bibinfo{address}{Oxford},
  \bibinfo{year}{2003}).

\bibitem[{\citenamefont{{Kevrekidis} et~al.}(2008)\citenamefont{{Kevrekidis},
  {Frantzeskakis}, and {Carretero-Gonz{\'a}lez}}}]{EmergentNL}
\bibinfo{author}{\bibfnamefont{P.~G.} \bibnamefont{{Kevrekidis}}},
  \bibinfo{author}{\bibfnamefont{D.~J.} \bibnamefont{{Frantzeskakis}}},
  \bibnamefont{and}
  \bibinfo{author}{\bibfnamefont{R.}~\bibnamefont{{Carretero-Gonz{\'a}lez}}},
  \emph{\bibinfo{title}{{Emergent Nonlinear Phenomena in Bose-Einstein
  Condensates}}} (\bibinfo{year}{2008}).

\bibitem[{\citenamefont{Bagnato et~al.}(2015)\citenamefont{Bagnato,
  Frantzeskakis, Kevrekidis, Malomed, and Mihalache}}]{LA-UR-15-20791}
\bibinfo{author}{\bibfnamefont{V.~S.} \bibnamefont{Bagnato}},
  \bibinfo{author}{\bibfnamefont{D.~J.} \bibnamefont{Frantzeskakis}},
  \bibinfo{author}{\bibfnamefont{P.~G.} \bibnamefont{Kevrekidis}},
  \bibinfo{author}{\bibfnamefont{B.~A.} \bibnamefont{Malomed}},
  \bibnamefont{and} \bibinfo{author}{\bibfnamefont{D.}~\bibnamefont{Mihalache}}
  (\bibinfo{year}{2015}), \bibinfo{note}{rom. Rep. Phys. {\bf 67}, 5}.

\bibitem[{\citenamefont{Cornell and Wieman}(2002)}]{RevModPhys.74.875}
\bibinfo{author}{\bibfnamefont{E.~A.} \bibnamefont{Cornell}} \bibnamefont{and}
  \bibinfo{author}{\bibfnamefont{C.~E.} \bibnamefont{Wieman}},
  \bibinfo{journal}{Rev. Mod. Phys.} \textbf{\bibinfo{volume}{74}},
  \bibinfo{pages}{875} (\bibinfo{year}{2002}).

\bibitem[{\citenamefont{Dalfovo et~al.}(1999)\citenamefont{Dalfovo, Giorgini,
  Pitaevskii, and Stringari}}]{RevModPhys.71.463}
\bibinfo{author}{\bibfnamefont{F.}~\bibnamefont{Dalfovo}},
  \bibinfo{author}{\bibfnamefont{S.}~\bibnamefont{Giorgini}},
  \bibinfo{author}{\bibfnamefont{L.~P.} \bibnamefont{Pitaevskii}},
  \bibnamefont{and}
  \bibinfo{author}{\bibfnamefont{S.}~\bibnamefont{Stringari}},
  \bibinfo{journal}{Rev. Mod. Phys.} \textbf{\bibinfo{volume}{71}},
  \bibinfo{pages}{463} (\bibinfo{year}{1999}).

\bibitem[{\citenamefont{Ketterle}(2002)}]{RevModPhys.74.1131}
\bibinfo{author}{\bibfnamefont{W.}~\bibnamefont{Ketterle}},
  \bibinfo{journal}{Rev. Mod. Phys.} \textbf{\bibinfo{volume}{74}},
  \bibinfo{pages}{1131} (\bibinfo{year}{2002}).

\bibitem[{\citenamefont{Leanhardt et~al.}(2002)\citenamefont{Leanhardt,
  Chikkatur, Kielpinski, Shin, Gustavson, Ketterle, and
  Pritchard}}]{PhysRevLett.89.040401}
\bibinfo{author}{\bibfnamefont{A.~E.} \bibnamefont{Leanhardt}},
  \bibinfo{author}{\bibfnamefont{A.~P.} \bibnamefont{Chikkatur}},
  \bibinfo{author}{\bibfnamefont{D.}~\bibnamefont{Kielpinski}},
  \bibinfo{author}{\bibfnamefont{Y.}~\bibnamefont{Shin}},
  \bibinfo{author}{\bibfnamefont{T.~L.} \bibnamefont{Gustavson}},
  \bibinfo{author}{\bibfnamefont{W.}~\bibnamefont{Ketterle}}, \bibnamefont{and}
  \bibinfo{author}{\bibfnamefont{D.~E.} \bibnamefont{Pritchard}},
  \bibinfo{journal}{Phys. Rev. Lett.} \textbf{\bibinfo{volume}{89}},
  \bibinfo{pages}{040401} (\bibinfo{year}{2002}).

\bibitem[{\citenamefont{Henderson et~al.}(2009)\citenamefont{Henderson, Ryu,
  MacCormick, and Boshier}}]{1367-2630-11-4-043030}
\bibinfo{author}{\bibfnamefont{K.}~\bibnamefont{Henderson}},
  \bibinfo{author}{\bibfnamefont{C.}~\bibnamefont{Ryu}},
  \bibinfo{author}{\bibfnamefont{C.}~\bibnamefont{MacCormick}},
  \bibnamefont{and} \bibinfo{author}{\bibfnamefont{M.~G.}
  \bibnamefont{Boshier}}, \bibinfo{journal}{New Journal of Physics}
  \textbf{\bibinfo{volume}{11}}, \bibinfo{pages}{043030}
  (\bibinfo{year}{2009}).

\bibitem[{\citenamefont{Gaunt and Hadzibabic}(2012)}]{dig_hol}
\bibinfo{author}{\bibfnamefont{A.~L.} \bibnamefont{Gaunt}} \bibnamefont{and}
  \bibinfo{author}{\bibfnamefont{Z.}~\bibnamefont{Hadzibabic}},
  \bibinfo{journal}{Scientific Reports} \textbf{\bibinfo{volume}{2}},
  \bibinfo{pages}{721} (\bibinfo{year}{2012}).

\bibitem[{\citenamefont{Pasienski and DeMarco}(2008)}]{Pasienski:08}
\bibinfo{author}{\bibfnamefont{M.}~\bibnamefont{Pasienski}} \bibnamefont{and}
  \bibinfo{author}{\bibfnamefont{B.}~\bibnamefont{DeMarco}},
  \bibinfo{journal}{Opt. Express} \textbf{\bibinfo{volume}{16}},
  \bibinfo{pages}{2176} (\bibinfo{year}{2008}).

\bibitem[{\citenamefont{Pollack et~al.}(2009)\citenamefont{Pollack, Dries,
  Junker, Chen, Corcovilos, and Hulet}}]{rhh}
\bibinfo{author}{\bibfnamefont{S.~E.} \bibnamefont{Pollack}},
  \bibinfo{author}{\bibfnamefont{D.}~\bibnamefont{Dries}},
  \bibinfo{author}{\bibfnamefont{M.}~\bibnamefont{Junker}},
  \bibinfo{author}{\bibfnamefont{Y.~P.} \bibnamefont{Chen}},
  \bibinfo{author}{\bibfnamefont{T.~A.} \bibnamefont{Corcovilos}},
  \bibnamefont{and} \bibinfo{author}{\bibfnamefont{R.~G.} \bibnamefont{Hulet}},
  \bibinfo{journal}{Phys. Rev. Lett.} \textbf{\bibinfo{volume}{102}},
  \bibinfo{pages}{090402} (\bibinfo{year}{2009}).
%  \urlprefix\url{https://link.aps.org/doi/10.1103/PhysRevLett.102.090402}.

\bibitem[{\citenamefont{Inouye et~al.}(1998)\citenamefont{Inouye, Andrews,
  Stenger, Miesner, Stamper-Kurn, and Ketterle}}]{Inouye1998Observation}
\bibinfo{author}{\bibfnamefont{S.}~\bibnamefont{Inouye}},
  \bibinfo{author}{\bibfnamefont{M.~R.} \bibnamefont{Andrews}},
  \bibinfo{author}{\bibfnamefont{J.}~\bibnamefont{Stenger}},
  \bibinfo{author}{\bibfnamefont{H.~J.} \bibnamefont{Miesner}},
  \bibinfo{author}{\bibfnamefont{D.~M.} \bibnamefont{Stamper-Kurn}},
  \bibnamefont{and} \bibinfo{author}{\bibfnamefont{W.}~\bibnamefont{Ketterle}},
  \bibinfo{journal}{Nature} \textbf{\bibinfo{volume}{392}},
  \bibinfo{pages}{151} (\bibinfo{year}{1998}).

\bibitem[{\citenamefont{{Frantzeskakis}}(2010)}]{2010JPhA.43u3001F}
\bibinfo{author}{\bibfnamefont{D.~J.} \bibnamefont{{Frantzeskakis}}},
  \bibinfo{journal}{Journal of Physics A Mathematical General}
  \textbf{\bibinfo{volume}{43}}, \bibinfo{eid}{213001} (\bibinfo{year}{2010}).

\bibitem[{\citenamefont{Strecker et~al.}(2003)\citenamefont{Strecker,
  Partridge, Truscott, and Hulet}}]{rhh2}
\bibinfo{author}{\bibfnamefont{K.~E.} \bibnamefont{Strecker}},
  \bibinfo{author}{\bibfnamefont{G.~B.} \bibnamefont{Partridge}},
  \bibinfo{author}{\bibfnamefont{A.~G.} \bibnamefont{Truscott}},
  \bibnamefont{and} \bibinfo{author}{\bibfnamefont{R.~G.} \bibnamefont{Hulet}},
  \bibinfo{journal}{New Journal of Physics} \textbf{\bibinfo{volume}{5}},
  \bibinfo{pages}{73} (\bibinfo{year}{2003}).
%  \urlprefix\url{http://stacks.iop.org/1367-2630/5/i=1/a=373}.

\bibitem[{\citenamefont{Kh.~Abdullaev et~al.}(2005)\citenamefont{Kh.~Abdullaev,
  Gammal, Kamchatnov, and Tomio}}]{tomio}
\bibinfo{author}{\bibfnamefont{F.}~\bibnamefont{Kh.~Abdullaev}},
  \bibinfo{author}{\bibfnamefont{A.}~\bibnamefont{Gammal}},
  \bibinfo{author}{\bibfnamefont{A.}~\bibnamefont{Kamchatnov}},
  \bibnamefont{and} \bibinfo{author}{\bibfnamefont{L.}~\bibnamefont{Tomio}},
  \bibinfo{journal}{International Journal of Modern Physics B}
  \textbf{\bibinfo{volume}{19}}, \bibinfo{pages}{3415} (\bibinfo{year}{2005}).
%  \eprint{https://doi.org/10.1142/S0217979205032279},
%  \urlprefix\url{https://doi.org/10.1142/S0217979205032279}.


\bibitem[{\citenamefont{Khaykovich et~al.}(2002)\citenamefont{Khaykovich,
  Schreck, Ferrari, Bourdel, Cubizolles, Carr, Castin, and
  Salomon}}]{Khaykovich1290}
\bibinfo{author}{\bibfnamefont{L.}~\bibnamefont{Khaykovich}},
  \bibinfo{author}{\bibfnamefont{F.}~\bibnamefont{Schreck}},
  \bibinfo{author}{\bibfnamefont{G.}~\bibnamefont{Ferrari}},
  \bibinfo{author}{\bibfnamefont{T.}~\bibnamefont{Bourdel}},
  \bibinfo{author}{\bibfnamefont{J.}~\bibnamefont{Cubizolles}},
  \bibinfo{author}{\bibfnamefont{L.~D.} \bibnamefont{Carr}},
  \bibinfo{author}{\bibfnamefont{Y.}~\bibnamefont{Castin}}, \bibnamefont{and}
  \bibinfo{author}{\bibfnamefont{C.}~\bibnamefont{Salomon}},
  \bibinfo{journal}{Science} \textbf{\bibinfo{volume}{296}},
  \bibinfo{pages}{1290} (\bibinfo{year}{2002}).

\bibitem[{\citenamefont{Strecker et~al.}(2002)\citenamefont{Strecker,
  Partridge, Truscott, and Hulet}}]{Strecker2002}
\bibinfo{author}{\bibfnamefont{K.}~\bibnamefont{Strecker}},
  \bibinfo{author}{\bibfnamefont{G.}~\bibnamefont{Partridge}},
  \bibinfo{author}{\bibfnamefont{A.~G.} \bibnamefont{Truscott}},
  \bibnamefont{and} \bibinfo{author}{\bibfnamefont{R.~H.} \bibnamefont{Hullet}},
  \bibinfo{journal}{Nature} \textbf{\bibinfo{volume}{417}},
  \bibinfo{pages}{150} (\bibinfo{year}{2002}).

\bibitem[{\citenamefont{Cornish et~al.}(2006)\citenamefont{Cornish, Thompson,
  and Wieman}}]{PhysRevLett.96.170401}
\bibinfo{author}{\bibfnamefont{S.~L.} \bibnamefont{Cornish}},
  \bibinfo{author}{\bibfnamefont{S.~T.} \bibnamefont{Thompson}},
  \bibnamefont{and} \bibinfo{author}{\bibfnamefont{C.~E.}
  \bibnamefont{Wieman}}, \bibinfo{journal}{Phys. Rev. Lett.}
  \textbf{\bibinfo{volume}{96}}, \bibinfo{pages}{170401}
  (\bibinfo{year}{2006}).

\bibitem[{\citenamefont{Burger et~al.}(1999)\citenamefont{Burger, Bongs,
  Dettmer, Ertmer, Sengstock, Sanpera, Shlyapnikov, and
  Lewenstein}}]{PhysRevLett.83.5198}
\bibinfo{author}{\bibfnamefont{S.}~\bibnamefont{Burger}},
  \bibinfo{author}{\bibfnamefont{K.}~\bibnamefont{Bongs}},
  \bibinfo{author}{\bibfnamefont{S.}~\bibnamefont{Dettmer}},
  \bibinfo{author}{\bibfnamefont{W.}~\bibnamefont{Ertmer}},
  \bibinfo{author}{\bibfnamefont{K.}~\bibnamefont{Sengstock}},
  \bibinfo{author}{\bibfnamefont{A.}~\bibnamefont{Sanpera}},
  \bibinfo{author}{\bibfnamefont{G.~V.} \bibnamefont{Shlyapnikov}},
  \bibnamefont{and}
  \bibinfo{author}{\bibfnamefont{M.}~\bibnamefont{Lewenstein}},
  \bibinfo{journal}{Phys. Rev. Lett.} \textbf{\bibinfo{volume}{83}},
  \bibinfo{pages}{5198} (\bibinfo{year}{1999}).

\bibitem[{\citenamefont{Anderson et~al.}(2001)\citenamefont{Anderson, Haljan,
  Regal, Feder, Collins, Clark, and Cornell}}]{PhysRevLett.86.2926}
\bibinfo{author}{\bibfnamefont{B.~P.} \bibnamefont{Anderson}},
  \bibinfo{author}{\bibfnamefont{P.~C.} \bibnamefont{Haljan}},
  \bibinfo{author}{\bibfnamefont{C.~A.} \bibnamefont{Regal}},
  \bibinfo{author}{\bibfnamefont{D.~L.} \bibnamefont{Feder}},
  \bibinfo{author}{\bibfnamefont{L.~A.} \bibnamefont{Collins}},
  \bibinfo{author}{\bibfnamefont{C.~W.} \bibnamefont{Clark}}, \bibnamefont{and}
  \bibinfo{author}{\bibfnamefont{E.~A.} \bibnamefont{Cornell}},
  \bibinfo{journal}{Phys. Rev. Lett.} \textbf{\bibinfo{volume}{86}},
  \bibinfo{pages}{2926} (\bibinfo{year}{2001}).

\bibitem[{\citenamefont{{Becker} et~al.}(2008)\citenamefont{{Becker},
  {Stellmer}, {Soltan-Panahi}, {D{\"o}rscher}, {Baumert}, {Richter},
  {Kronj{\"a}ger}, {Bongs}, and {Sengstock}}}]{2008NatPh.4.496B}
\bibinfo{author}{\bibfnamefont{C.}~\bibnamefont{{Becker}}},
  \bibinfo{author}{\bibfnamefont{S.}~\bibnamefont{{Stellmer}}},
  \bibinfo{author}{\bibfnamefont{P.}~\bibnamefont{{Soltan-Panahi}}},
  \bibinfo{author}{\bibfnamefont{S.}~\bibnamefont{{D{\"o}rscher}}},
  \bibinfo{author}{\bibfnamefont{M.}~\bibnamefont{{Baumert}}},
  \bibinfo{author}{\bibfnamefont{E.-M.} \bibnamefont{{Richter}}},
  \bibinfo{author}{\bibfnamefont{J.}~\bibnamefont{{Kronj{\"a}ger}}},
  \bibinfo{author}{\bibfnamefont{K.}~\bibnamefont{{Bongs}}}, \bibnamefont{and}
  \bibinfo{author}{\bibfnamefont{K.}~\bibnamefont{{Sengstock}}},
  \bibinfo{journal}{Nature Physics} \textbf{\bibinfo{volume}{4}},
  \bibinfo{pages}{496} (\bibinfo{year}{2008}).

\bibitem[{\citenamefont{Weller et~al.}(2008)\citenamefont{Weller, Ronzheimer,
  Gross, Esteve, Oberthaler, Frantzeskakis, Theocharis, and
  Kevrekidis}}]{markus}
\bibinfo{author}{\bibfnamefont{A.}~\bibnamefont{Weller}},
  \bibinfo{author}{\bibfnamefont{J.~P.} \bibnamefont{Ronzheimer}},
  \bibinfo{author}{\bibfnamefont{C.}~\bibnamefont{Gross}},
  \bibinfo{author}{\bibfnamefont{J.}~\bibnamefont{Esteve}},
  \bibinfo{author}{\bibfnamefont{M.~K.} \bibnamefont{Oberthaler}},
  \bibinfo{author}{\bibfnamefont{D.~J.} \bibnamefont{Frantzeskakis}},
  \bibinfo{author}{\bibfnamefont{G.}~\bibnamefont{Theocharis}},
  \bibnamefont{and} \bibinfo{author}{\bibfnamefont{P.~G.}
  \bibnamefont{Kevrekidis}}, \bibinfo{journal}{Phys. Rev. Lett.}
  \textbf{\bibinfo{volume}{101}}, \bibinfo{pages}{130401}
  (\bibinfo{year}{2008}).
%  \urlprefix\url{https://link.aps.org/doi/10.1103/PhysRevLett.101.130401}.

\bibitem[{\citenamefont{Denschlag et~al.}(2000)\citenamefont{Denschlag,
  Simsarian, Feder, Clark, Collins, Cubizolles, Deng, Hagley, Helmerson,
  Reinhardt et~al.}}]{Denschlag97}
\bibinfo{author}{\bibfnamefont{J.}~\bibnamefont{Denschlag}},
  \bibinfo{author}{\bibfnamefont{J.~E.} \bibnamefont{Simsarian}},
  \bibinfo{author}{\bibfnamefont{D.~L.} \bibnamefont{Feder}},
  \bibinfo{author}{\bibfnamefont{C.~W.} \bibnamefont{Clark}},
  \bibinfo{author}{\bibfnamefont{L.~A.} \bibnamefont{Collins}},
  \bibinfo{author}{\bibfnamefont{J.}~\bibnamefont{Cubizolles}},
  \bibinfo{author}{\bibfnamefont{L.}~\bibnamefont{Deng}},
  \bibinfo{author}{\bibfnamefont{E.~W.} \bibnamefont{Hagley}},
  \bibinfo{author}{\bibfnamefont{K.}~\bibnamefont{Helmerson}},
  \bibinfo{author}{\bibfnamefont{W.~P.} \bibnamefont{Reinhardt}},
  \bibnamefont{et~al.}, \bibinfo{journal}{Science}
  \textbf{\bibinfo{volume}{287}}, \bibinfo{pages}{97} (\bibinfo{year}{2000}).

\bibitem[{\citenamefont{Dutton et~al.}(2001)\citenamefont{Dutton, Budde, Slowe,
  and Hau}}]{Dutton663}
\bibinfo{author}{\bibfnamefont{Z.}~\bibnamefont{Dutton}},
  \bibinfo{author}{\bibfnamefont{M.}~\bibnamefont{Budde}},
  \bibinfo{author}{\bibfnamefont{C.}~\bibnamefont{Slowe}}, \bibnamefont{and}
  \bibinfo{author}{\bibfnamefont{L.~V.} \bibnamefont{Hau}},
  \bibinfo{journal}{Science} \textbf{\bibinfo{volume}{293}},
  \bibinfo{pages}{663} (\bibinfo{year}{2001}).

\bibitem[{\citenamefont{Hamner et~al.}(2011)\citenamefont{Hamner, Chang,
  Engels, and Hoefer}}]{PhysRevLett.106.065302}
\bibinfo{author}{\bibfnamefont{C.}~\bibnamefont{Hamner}},
  \bibinfo{author}{\bibfnamefont{J.~J.} \bibnamefont{Chang}},
  \bibinfo{author}{\bibfnamefont{P.}~\bibnamefont{Engels}}, \bibnamefont{and}
  \bibinfo{author}{\bibfnamefont{M.~A.} \bibnamefont{Hoefer}},
  \bibinfo{journal}{Phys. Rev. Lett.} \textbf{\bibinfo{volume}{106}},
  \bibinfo{pages}{065302} (\bibinfo{year}{2011}).

\bibitem[{\citenamefont{Middelkamp
  et~al.}(2011{\natexlab{a}})\citenamefont{Middelkamp, Chang, Hamner,
  Carretero-Gonz\'alez, Kevrekidis, Achilleos, Frantzeskakis, Schmelcher, and
  Engels}}]{MIDDELKAMP2011642}
\bibinfo{author}{\bibfnamefont{S.}~\bibnamefont{Middelkamp}},
  \bibinfo{author}{\bibfnamefont{J.}~\bibnamefont{Chang}},
  \bibinfo{author}{\bibfnamefont{C.}~\bibnamefont{Hamner}},
  \bibinfo{author}{\bibfnamefont{R.}~\bibnamefont{Carretero-Gonz\'alez}},
  \bibinfo{author}{\bibfnamefont{P.}~\bibnamefont{Kevrekidis}},
  \bibinfo{author}{\bibfnamefont{V.}~\bibnamefont{Achilleos}},
  \bibinfo{author}{\bibfnamefont{D.}~\bibnamefont{Frantzeskakis}},
  \bibinfo{author}{\bibfnamefont{P.}~\bibnamefont{Schmelcher}},
  \bibnamefont{and} \bibinfo{author}{\bibfnamefont{P.}~\bibnamefont{Engels}},
  \bibinfo{journal}{Physics Letters A} \textbf{\bibinfo{volume}{375}},
  \bibinfo{pages}{642 } (\bibinfo{year}{2011}{\natexlab{a}}).

\bibitem[{\citenamefont{Eiermann et~al.}(2004)\citenamefont{Eiermann, Anker,
  Albiez, Taglieber, Treutlein, Marzlin, and Oberthaler}}]{gap}
\bibinfo{author}{\bibfnamefont{B.}~\bibnamefont{Eiermann}},
  \bibinfo{author}{\bibfnamefont{T.}~\bibnamefont{Anker}},
  \bibinfo{author}{\bibfnamefont{M.}~\bibnamefont{Albiez}},
  \bibinfo{author}{\bibfnamefont{M.}~\bibnamefont{Taglieber}},
  \bibinfo{author}{\bibfnamefont{P.}~\bibnamefont{Treutlein}},
  \bibinfo{author}{\bibfnamefont{K.-P.} \bibnamefont{Marzlin}},
  \bibnamefont{and} \bibinfo{author}{\bibfnamefont{M.~K.}
  \bibnamefont{Oberthaler}}, \bibinfo{journal}{Phys. Rev. Lett.}
  \textbf{\bibinfo{volume}{92}}, \bibinfo{pages}{230401}
  (\bibinfo{year}{2004}).
%  \urlprefix\url{https://link.aps.org/doi/10.1103/PhysRevLett.92.230401}.

\bibitem[{\citenamefont{Morsch and Oberthaler}(2006)}]{RevModPhys.78.179}
\bibinfo{author}{\bibfnamefont{O.}~\bibnamefont{Morsch}} \bibnamefont{and}
  \bibinfo{author}{\bibfnamefont{M.}~\bibnamefont{Oberthaler}},
  \bibinfo{journal}{Rev. Mod. Phys.} \textbf{\bibinfo{volume}{78}},
  \bibinfo{pages}{179} (\bibinfo{year}{2006}).

\bibitem[{\citenamefont{Kevrekidis et~al.}(2015)\citenamefont{Kevrekidis,
  Frantzeskakis, and Carretero-Gonz\'alez}}]{doi:10.1137/1.9781611973945}
\bibinfo{author}{\bibfnamefont{P.}~\bibnamefont{Kevrekidis}},
  \bibinfo{author}{\bibfnamefont{D.}~\bibnamefont{Frantzeskakis}},
  \bibnamefont{and}
  \bibinfo{author}{\bibfnamefont{R.}~\bibnamefont{Carretero-Gonz\'alez}},
  \emph{\bibinfo{title}{The Defocusing Nonlinear Schr\"odinger Equation}}
  (\bibinfo{publisher}{Society for Industrial and Applied Mathematics},
  \bibinfo{address}{Philadelphia, PA}, \bibinfo{year}{2015}).

\bibitem[{\citenamefont{{Ablowitz} et~al.}(2004)\citenamefont{{Ablowitz},
  {Prinari}, and {Trubatch}}}]{2004dcns.bookA}
\bibinfo{author}{\bibfnamefont{M.~J.} \bibnamefont{{Ablowitz}}},
  \bibinfo{author}{\bibfnamefont{B.}~\bibnamefont{{Prinari}}},
  \bibnamefont{and} \bibinfo{author}{\bibfnamefont{A.~D.}
  \bibnamefont{{Trubatch}}}, \emph{\bibinfo{title}{{Discrete and Continuous
  Nonlinear Schr{\"o}dinger Systems}}} (\bibinfo{year}{2004}).

\bibitem[{\citenamefont{Sulem and Sulem}(2007)}]{sulem2007nonlinear}
\bibinfo{author}{\bibfnamefont{C.}~\bibnamefont{Sulem}} \bibnamefont{and}
  \bibinfo{author}{\bibfnamefont{P.}~\bibnamefont{Sulem}},
  \emph{\bibinfo{title}{The Nonlinear Schr{\"o}dinger Equation: Self-Focusing
  and Wave Collapse}}, Applied Mathematical Sciences
  (\bibinfo{publisher}{Springer New York}, \bibinfo{year}{2007}).

\bibitem[{\citenamefont{Kevrekidis and
  Frantzeskakis}(2004)}]{doi:10.1142/S0217984904006809}
\bibinfo{author}{\bibfnamefont{P.~G.} \bibnamefont{Kevrekidis}}
  \bibnamefont{and} \bibinfo{author}{\bibfnamefont{D.~J.}
  \bibnamefont{Frantzeskakis}}, \bibinfo{journal}{Modern Physics Letters B}
  \textbf{\bibinfo{volume}{18}}, \bibinfo{pages}{173} (\bibinfo{year}{2004}).

\bibitem[{\citenamefont{M\"ott\"onen et~al.}(2005)\citenamefont{M\"ott\"onen,
  Virtanen, Isoshima, and Salomaa}}]{PhysRevA.71.033626}
\bibinfo{author}{\bibfnamefont{M.}~\bibnamefont{M\"ott\"onen}},
  \bibinfo{author}{\bibfnamefont{S.~M.~M.} \bibnamefont{Virtanen}},
  \bibinfo{author}{\bibfnamefont{T.}~\bibnamefont{Isoshima}}, \bibnamefont{and}
  \bibinfo{author}{\bibfnamefont{M.~M.} \bibnamefont{Salomaa}},
  \bibinfo{journal}{Phys. Rev. A} \textbf{\bibinfo{volume}{71}},
  \bibinfo{pages}{033626} (\bibinfo{year}{2005}).

\bibitem[{\citenamefont{Pietil\"a et~al.}(2006)\citenamefont{Pietil\"a,
  M\"ott\"onen, Isoshima, Huhtam\"aki, and Virtanen}}]{PhysRevA.74.023603}
\bibinfo{author}{\bibfnamefont{V.}~\bibnamefont{Pietil\"a}},
  \bibinfo{author}{\bibfnamefont{M.}~\bibnamefont{M\"ott\"onen}},
  \bibinfo{author}{\bibfnamefont{T.}~\bibnamefont{Isoshima}},
  \bibinfo{author}{\bibfnamefont{J.~A.~M.} \bibnamefont{Huhtam\"aki}},
  \bibnamefont{and} \bibinfo{author}{\bibfnamefont{S.~M.~M.}
  \bibnamefont{Virtanen}}, \bibinfo{journal}{Phys. Rev. A}
  \textbf{\bibinfo{volume}{74}}, \bibinfo{pages}{023603}
  (\bibinfo{year}{2006}).

\bibitem[{\citenamefont{Middelkamp et~al.}(2010)\citenamefont{Middelkamp,
  Kevrekidis, Frantzeskakis, Carretero-Gonz\'alez, and
  Schmelcher}}]{PhysRevA.82.013646}
\bibinfo{author}{\bibfnamefont{S.}~\bibnamefont{Middelkamp}},
  \bibinfo{author}{\bibfnamefont{P.~G.} \bibnamefont{Kevrekidis}},
  \bibinfo{author}{\bibfnamefont{D.~J.} \bibnamefont{Frantzeskakis}},
  \bibinfo{author}{\bibfnamefont{R.}~\bibnamefont{Carretero-Gonz\'alez}},
  \bibnamefont{and}
  \bibinfo{author}{\bibfnamefont{P.}~\bibnamefont{Schmelcher}},
  \bibinfo{journal}{Phys. Rev. A} \textbf{\bibinfo{volume}{82}},
  \bibinfo{pages}{013646} (\bibinfo{year}{2010}).

\bibitem[{\citenamefont{Middelkamp
  et~al.}(2011{\natexlab{b}})\citenamefont{Middelkamp, Kevrekidis,
  Frantzeskakis, Carretero-Gonz\'alez, and Schmelcher}}]{MIDDELKAMP20111449}
\bibinfo{author}{\bibfnamefont{S.}~\bibnamefont{Middelkamp}},
  \bibinfo{author}{\bibfnamefont{P.}~\bibnamefont{Kevrekidis}},
  \bibinfo{author}{\bibfnamefont{D.}~\bibnamefont{Frantzeskakis}},
  \bibinfo{author}{\bibfnamefont{R.}~\bibnamefont{Carretero-Gonz\'alez}},
  \bibnamefont{and}
  \bibinfo{author}{\bibfnamefont{P.}~\bibnamefont{Schmelcher}},
  \bibinfo{journal}{Physica D: Nonlinear Phenomena}
  \textbf{\bibinfo{volume}{240}}, \bibinfo{pages}{1449 }
  (\bibinfo{year}{2011}{\natexlab{b}}).

\bibitem[{\citenamefont{Milburn et~al.}(1997)\citenamefont{Milburn, Corney,
  Wright, and Walls}}]{PhysRevA.55.4318}
\bibinfo{author}{\bibfnamefont{G.~J.} \bibnamefont{Milburn}},
  \bibinfo{author}{\bibfnamefont{J.}~\bibnamefont{Corney}},
  \bibinfo{author}{\bibfnamefont{E.~M.} \bibnamefont{Wright}},
  \bibnamefont{and} \bibinfo{author}{\bibfnamefont{D.~F.} \bibnamefont{Walls}},
  \bibinfo{journal}{Phys. Rev. A} \textbf{\bibinfo{volume}{55}},
  \bibinfo{pages}{4318} (\bibinfo{year}{1997}).

\bibitem[{\citenamefont{Capuzzi and Hern\'andez}(1999)}]{PhysRevA.59.1488}
\bibinfo{author}{\bibfnamefont{P.}~\bibnamefont{Capuzzi}} \bibnamefont{and}
  \bibinfo{author}{\bibfnamefont{E.~S.} \bibnamefont{Hern\'andez}},
  \bibinfo{journal}{Phys. Rev. A} \textbf{\bibinfo{volume}{59}},
  \bibinfo{pages}{1488} (\bibinfo{year}{1999}).

\bibitem[{\citenamefont{Holthaus}(2001)}]{PhysRevA.64.011601}
\bibinfo{author}{\bibfnamefont{M.}~\bibnamefont{Holthaus}},
  \bibinfo{journal}{Phys. Rev. A} \textbf{\bibinfo{volume}{64}},
  \bibinfo{pages}{011601} (\bibinfo{year}{2001}).

\bibitem[{\citenamefont{Shin et~al.}(2005)\citenamefont{Shin, Sanner, Jo,
  Pasquini, Saba, Ketterle, Pritchard, Vengalattore, and
  Prentiss}}]{PhysRevA.72.021604}
\bibinfo{author}{\bibfnamefont{Y.}~\bibnamefont{Shin}},
  \bibinfo{author}{\bibfnamefont{C.}~\bibnamefont{Sanner}},
  \bibinfo{author}{\bibfnamefont{G.-B.} \bibnamefont{Jo}},
  \bibinfo{author}{\bibfnamefont{T.~A.} \bibnamefont{Pasquini}},
  \bibinfo{author}{\bibfnamefont{M.}~\bibnamefont{Saba}},
  \bibinfo{author}{\bibfnamefont{W.}~\bibnamefont{Ketterle}},
  \bibinfo{author}{\bibfnamefont{D.~E.} \bibnamefont{Pritchard}},
  \bibinfo{author}{\bibfnamefont{M.}~\bibnamefont{Vengalattore}},
  \bibnamefont{and} \bibinfo{author}{\bibfnamefont{M.}~\bibnamefont{Prentiss}},
  \bibinfo{journal}{Phys. Rev. A} \textbf{\bibinfo{volume}{72}},
  \bibinfo{pages}{021604} (\bibinfo{year}{2005}).

\bibitem[{\citenamefont{Wang et~al.}(2005)\citenamefont{Wang, Anderson, Bright,
  Cornell, Diot, Kishimoto, Prentiss, Saravanan, Segal, and
  Wu}}]{PhysRevLett.94.090405}
\bibinfo{author}{\bibfnamefont{Y.-J.} \bibnamefont{Wang}},
  \bibinfo{author}{\bibfnamefont{D.~Z.} \bibnamefont{Anderson}},
  \bibinfo{author}{\bibfnamefont{V.~M.} \bibnamefont{Bright}},
  \bibinfo{author}{\bibfnamefont{E.~A.} \bibnamefont{Cornell}},
  \bibinfo{author}{\bibfnamefont{Q.}~\bibnamefont{Diot}},
  \bibinfo{author}{\bibfnamefont{T.}~\bibnamefont{Kishimoto}},
  \bibinfo{author}{\bibfnamefont{M.}~\bibnamefont{Prentiss}},
  \bibinfo{author}{\bibfnamefont{R.~A.} \bibnamefont{Saravanan}},
  \bibinfo{author}{\bibfnamefont{S.~R.} \bibnamefont{Segal}}, \bibnamefont{and}
  \bibinfo{author}{\bibfnamefont{S.}~\bibnamefont{Wu}}, \bibinfo{journal}{Phys.
  Rev. Lett.} \textbf{\bibinfo{volume}{94}}, \bibinfo{pages}{090405}
  (\bibinfo{year}{2005}).

\bibitem[{\citenamefont{Albiez et~al.}(2005)\citenamefont{Albiez, Gati,
  F\"olling, Hunsmann, Cristiani, and Oberthaler}}]{PhysRevLett.95.010402}
\bibinfo{author}{\bibfnamefont{M.}~\bibnamefont{Albiez}},
  \bibinfo{author}{\bibfnamefont{R.}~\bibnamefont{Gati}},
  \bibinfo{author}{\bibfnamefont{J.}~\bibnamefont{F\"olling}},
  \bibinfo{author}{\bibfnamefont{S.}~\bibnamefont{Hunsmann}},
  \bibinfo{author}{\bibfnamefont{M.}~\bibnamefont{Cristiani}},
  \bibnamefont{and} \bibinfo{author}{\bibfnamefont{M.~K.}
  \bibnamefont{Oberthaler}}, \bibinfo{journal}{Phys. Rev. Lett.}
  \textbf{\bibinfo{volume}{95}}, \bibinfo{pages}{010402}
  (\bibinfo{year}{2005}).

\bibitem[{\citenamefont{Zibold et~al.}(2010)\citenamefont{Zibold, Nicklas,
  Gross, and Oberthaler}}]{tilman}
\bibinfo{author}{\bibfnamefont{T.}~\bibnamefont{Zibold}},
  \bibinfo{author}{\bibfnamefont{E.}~\bibnamefont{Nicklas}},
  \bibinfo{author}{\bibfnamefont{C.}~\bibnamefont{Gross}}, \bibnamefont{and}
  \bibinfo{author}{\bibfnamefont{M.~K.} \bibnamefont{Oberthaler}},
  \bibinfo{journal}{Phys. Rev. Lett.} \textbf{\bibinfo{volume}{105}},
  \bibinfo{pages}{204101} (\bibinfo{year}{2010}).
%  \urlprefix\url{https://link.aps.org/doi/10.1103/PhysRevLett.105.204101}.

\bibitem[{\citenamefont{{Wang} et~al.}(2009)\citenamefont{{Wang}, {Theocharis},
  {Kevrekidis}, {Whitaker}, {Law}, {Frantzeskakis}, and
  {Malomed}}}]{2009PhRvE.80d6611W}
\bibinfo{author}{\bibfnamefont{C.}~\bibnamefont{{Wang}}},
  \bibinfo{author}{\bibfnamefont{G.}~\bibnamefont{{Theocharis}}},
  \bibinfo{author}{\bibfnamefont{P.~G.} \bibnamefont{{Kevrekidis}}},
  \bibinfo{author}{\bibfnamefont{N.}~\bibnamefont{{Whitaker}}},
  \bibinfo{author}{\bibfnamefont{K.~J.~H.} \bibnamefont{{Law}}},
  \bibinfo{author}{\bibfnamefont{D.~J.} \bibnamefont{{Frantzeskakis}}},
  \bibnamefont{and} \bibinfo{author}{\bibfnamefont{B.~A.}
  \bibnamefont{{Malomed}}}, \bibinfo{journal}{\pre}
  \textbf{\bibinfo{volume}{80}}, \bibinfo{eid}{046611} (\bibinfo{year}{2009}),
  \eprint{arXiv:0904.0255}.

\bibitem[{\citenamefont{Adhikari and Muruganandam}(2003)}]{ADHIKARI2003229}
\bibinfo{author}{\bibfnamefont{S.~K.} \bibnamefont{Adhikari}} \bibnamefont{and}
  \bibinfo{author}{\bibfnamefont{P.}~\bibnamefont{Muruganandam}},
  \bibinfo{journal}{Physics Letters A} \textbf{\bibinfo{volume}{310}},
  \bibinfo{pages}{229 } (\bibinfo{year}{2003}), ISSN \bibinfo{issn}{0375-9601}.

\bibitem[{\citenamefont{Choi and Niu}(1999)}]{PhysRevLett.82.2022}
\bibinfo{author}{\bibfnamefont{D.-I.} \bibnamefont{Choi}} \bibnamefont{and}
  \bibinfo{author}{\bibfnamefont{Q.}~\bibnamefont{Niu}},
  \bibinfo{journal}{Phys. Rev. Lett.} \textbf{\bibinfo{volume}{82}},
  \bibinfo{pages}{2022} (\bibinfo{year}{1999}).

\bibitem[{\citenamefont{Lin et~al.}(2009)\citenamefont{Lin, Perry, Compton,
  Spielman, and Porto}}]{PhysRevA.79.063631}
\bibinfo{author}{\bibfnamefont{Y.-J.} \bibnamefont{Lin}},
  \bibinfo{author}{\bibfnamefont{A.~R.} \bibnamefont{Perry}},
  \bibinfo{author}{\bibfnamefont{R.~L.} \bibnamefont{Compton}},
  \bibinfo{author}{\bibfnamefont{I.~B.} \bibnamefont{Spielman}},
  \bibnamefont{and} \bibinfo{author}{\bibfnamefont{J.~V.} \bibnamefont{Porto}},
  \bibinfo{journal}{Phys. Rev. A} \textbf{\bibinfo{volume}{79}},
  \bibinfo{pages}{063631} (\bibinfo{year}{2009}).

\bibitem[{\citenamefont{Pepino et~al.}(2009)\citenamefont{Pepino, Cooper,
  Anderson, and Holland}}]{PhysRevLett.103.140405}
\bibinfo{author}{\bibfnamefont{R.~A.} \bibnamefont{Pepino}},
  \bibinfo{author}{\bibfnamefont{J.}~\bibnamefont{Cooper}},
  \bibinfo{author}{\bibfnamefont{D.~Z.} \bibnamefont{Anderson}},
  \bibnamefont{and} \bibinfo{author}{\bibfnamefont{M.~J.}
  \bibnamefont{Holland}}, \bibinfo{journal}{Phys. Rev. Lett.}
  \textbf{\bibinfo{volume}{103}}, \bibinfo{pages}{140405}
  (\bibinfo{year}{2009}).

\bibitem[{\citenamefont{Seaman et~al.}(2007)\citenamefont{Seaman, Kr\"amer,
  Anderson, and Holland}}]{PhysRevA.75.023615}
\bibinfo{author}{\bibfnamefont{B.~T.} \bibnamefont{Seaman}},
  \bibinfo{author}{\bibfnamefont{M.}~\bibnamefont{Kr\"amer}},
  \bibinfo{author}{\bibfnamefont{D.~Z.} \bibnamefont{Anderson}},
  \bibnamefont{and} \bibinfo{author}{\bibfnamefont{M.~J.}
  \bibnamefont{Holland}}, \bibinfo{journal}{Phys. Rev. A}
  \textbf{\bibinfo{volume}{75}}, \bibinfo{pages}{023615}
  (\bibinfo{year}{2007}).

\bibitem[{\citenamefont{Ramanathan et~al.}(2011)\citenamefont{Ramanathan,
  Wright, Muniz, Zelan, Hill, Lobb, Helmerson, Phillips, and
  Campbell}}]{PhysRevLett.106.130401}
\bibinfo{author}{\bibfnamefont{A.}~\bibnamefont{Ramanathan}},
  \bibinfo{author}{\bibfnamefont{K.~C.} \bibnamefont{Wright}},
  \bibinfo{author}{\bibfnamefont{S.~R.} \bibnamefont{Muniz}},
  \bibinfo{author}{\bibfnamefont{M.}~\bibnamefont{Zelan}},
  \bibinfo{author}{\bibfnamefont{W.~T.} \bibnamefont{Hill}},
  \bibinfo{author}{\bibfnamefont{C.~J.} \bibnamefont{Lobb}},
  \bibinfo{author}{\bibfnamefont{K.}~\bibnamefont{Helmerson}},
  \bibinfo{author}{\bibfnamefont{W.~D.} \bibnamefont{Phillips}},
  \bibnamefont{and} \bibinfo{author}{\bibfnamefont{G.~K.}
  \bibnamefont{Campbell}}, \bibinfo{journal}{Phys. Rev. Lett.}
  \textbf{\bibinfo{volume}{106}}, \bibinfo{pages}{130401}
  (\bibinfo{year}{2011}).

\bibitem[{\citenamefont{Wright et~al.}(2013)\citenamefont{Wright, Blakestad,
  Lobb, Phillips, and Campbell}}]{PhysRevLett.110.025302}
\bibinfo{author}{\bibfnamefont{K.~C.} \bibnamefont{Wright}},
  \bibinfo{author}{\bibfnamefont{R.~B.} \bibnamefont{Blakestad}},
  \bibinfo{author}{\bibfnamefont{C.~J.} \bibnamefont{Lobb}},
  \bibinfo{author}{\bibfnamefont{W.~D.} \bibnamefont{Phillips}},
  \bibnamefont{and} \bibinfo{author}{\bibfnamefont{G.~K.}
  \bibnamefont{Campbell}}, \bibinfo{journal}{Phys. Rev. Lett.}
  \textbf{\bibinfo{volume}{110}}, \bibinfo{pages}{025302}
  (\bibinfo{year}{2013}).

\bibitem[{\citenamefont{Murray et~al.}(2013)\citenamefont{Murray, Krygier,
  Edwards, Wright, Campbell, and Clark}}]{PhysRevA.88.053615}
\bibinfo{author}{\bibfnamefont{N.}~\bibnamefont{Murray}},
  \bibinfo{author}{\bibfnamefont{M.}~\bibnamefont{Krygier}},
  \bibinfo{author}{\bibfnamefont{M.}~\bibnamefont{Edwards}},
  \bibinfo{author}{\bibfnamefont{K.~C.} \bibnamefont{Wright}},
  \bibinfo{author}{\bibfnamefont{G.~K.} \bibnamefont{Campbell}},
  \bibnamefont{and} \bibinfo{author}{\bibfnamefont{C.~W.} \bibnamefont{Clark}},
  \bibinfo{journal}{Phys. Rev. A} \textbf{\bibinfo{volume}{88}},
  \bibinfo{pages}{053615} (\bibinfo{year}{2013}).

\bibitem[{\citenamefont{Clarke and Braginski}(2004)}]{BraginskiBook}
\bibinfo{author}{\bibfnamefont{J.}~\bibnamefont{Clarke}} \bibnamefont{and}
  \bibinfo{author}{\bibfnamefont{A.~I.} \bibnamefont{Braginski}},
  \emph{\bibinfo{title}{{The SQUID Handbook: Fundamentals and Technology of
  SQUIDs and SQUID Systems}}} (\bibinfo{publisher}{Wiley-VCH},
  \bibinfo{year}{2004}), \bibinfo{edition}{1st} ed., ISBN
  \bibinfo{isbn}{3527402292}.

\bibitem[{\citenamefont{Eckel et~al.}(2017)\citenamefont{Eckel, Kumar,
  Jacobson, Spielman, and Campbell}}]{Eckel:2017uqx}
\bibinfo{author}{\bibfnamefont{S.}~\bibnamefont{Eckel}},
  \bibinfo{author}{\bibfnamefont{A.}~\bibnamefont{Kumar}},
  \bibinfo{author}{\bibfnamefont{T.}~\bibnamefont{Jacobson}},
  \bibinfo{author}{\bibfnamefont{I.~B.} \bibnamefont{Spielman}},
  \bibnamefont{and} \bibinfo{author}{\bibfnamefont{G.~K.}
  \bibnamefont{Campbell}} (\bibinfo{year}{2017}), \eprint{arXiv:1710.05800}.

\bibitem[{\citenamefont{Mathew et~al.}(2015)\citenamefont{Mathew, Kumar, Eckel,
  Jendrzejewski, Campbell, Edwards, and Tiesinga}}]{PhysRevA.92.033602}
\bibinfo{author}{\bibfnamefont{R.}~\bibnamefont{Mathew}},
  \bibinfo{author}{\bibfnamefont{A.}~\bibnamefont{Kumar}},
  \bibinfo{author}{\bibfnamefont{S.}~\bibnamefont{Eckel}},
  \bibinfo{author}{\bibfnamefont{F.}~\bibnamefont{Jendrzejewski}},
  \bibinfo{author}{\bibfnamefont{G.~K.} \bibnamefont{Campbell}},
  \bibinfo{author}{\bibfnamefont{M.}~\bibnamefont{Edwards}}, \bibnamefont{and}
  \bibinfo{author}{\bibfnamefont{E.}~\bibnamefont{Tiesinga}},
  \bibinfo{journal}{Phys. Rev. A} \textbf{\bibinfo{volume}{92}},
  \bibinfo{pages}{033602} (\bibinfo{year}{2015}).

\bibitem[{\citenamefont{Eckel et~al.}(2014)\citenamefont{Eckel, Lobb, Edwards,
  Phillips, Lee, Jendrzejewski, Murray, and Campbell}}]{eckel_2013}
\bibinfo{author}{\bibfnamefont{S.}~\bibnamefont{Eckel}},
  \bibinfo{author}{\bibfnamefont{J.}~\bibnamefont{Lobb}},
  \bibinfo{author}{\bibfnamefont{M.}~\bibnamefont{Edwards}},
  \bibinfo{author}{\bibfnamefont{W.}~\bibnamefont{Phillips}},
  \bibinfo{author}{\bibfnamefont{J.}~\bibnamefont{Lee}},
  \bibinfo{author}{\bibfnamefont{F.}~\bibnamefont{Jendrzejewski}},
  \bibinfo{author}{\bibfnamefont{N.}~\bibnamefont{Murray}}, \bibnamefont{and}
  \bibinfo{author}{\bibfnamefont{G.}~\bibnamefont{Campbell}},
  \bibinfo{journal}{Nature} \textbf{\bibinfo{volume}{506}},
  \bibinfo{pages}{200} (\bibinfo{year}{2014}).

\bibitem[{\citenamefont{Charalampidis et~al.}(2018)\citenamefont{Charalampidis,
  Kevrekidis, and Farrell}}]{CHARALAMPIDIS2018482}
\bibinfo{author}{\bibfnamefont{E.}~\bibnamefont{Charalampidis}},
  \bibinfo{author}{\bibfnamefont{P.}~\bibnamefont{Kevrekidis}},
  \bibnamefont{and} \bibinfo{author}{\bibfnamefont{P.}~\bibnamefont{Farrell}},
  \bibinfo{journal}{Communications in Nonlinear Science and Numerical
  Simulation} \textbf{\bibinfo{volume}{54}}, \bibinfo{pages}{482 }
  (\bibinfo{year}{2018}).

\bibitem[{\citenamefont{Stewart}(2002)}]{doi:10.1137/S0895479800371529}
\bibinfo{author}{\bibfnamefont{G.~W.} \bibnamefont{Stewart}},
  \bibinfo{journal}{SIAM Journal on Matrix Analysis and Applications}
  \textbf{\bibinfo{volume}{23}}, \bibinfo{pages}{601} (\bibinfo{year}{2002}).

\bibitem[{\citenamefont{Lehoucq et~al.}(1998)\citenamefont{Lehoucq, Sorensen,
  and Yang}}]{doi:10.1137/1.9780898719628}
\bibinfo{author}{\bibfnamefont{R.}~\bibnamefont{Lehoucq}},
  \bibinfo{author}{\bibfnamefont{D.}~\bibnamefont{Sorensen}}, \bibnamefont{and}
  \bibinfo{author}{\bibfnamefont{C.}~\bibnamefont{Yang}},
  \emph{\bibinfo{title}{ARPACK Users' Guide}} (\bibinfo{publisher}{Society for
  Industrial and Applied Mathematics}, \bibinfo{year}{1998}).

\bibitem[{\citenamefont{Hernandez et~al.}(2005)\citenamefont{Hernandez, Roman,
  and Vidal}}]{Hernandez:2005:SSF}
\bibinfo{author}{\bibfnamefont{V.}~\bibnamefont{Hernandez}},
  \bibinfo{author}{\bibfnamefont{J.~E.} \bibnamefont{Roman}}, \bibnamefont{and}
  \bibinfo{author}{\bibfnamefont{V.}~\bibnamefont{Vidal}},
  \bibinfo{journal}{{ACM} Trans. Math. Software} \textbf{\bibinfo{volume}{31}},
  \bibinfo{pages}{351} (\bibinfo{year}{2005}).

\bibitem[{\citenamefont{Hernandez et~al.}(2003)\citenamefont{Hernandez, Roman,
  and Vidal}}]{Hernandez:2003:SSL}
\bibinfo{author}{\bibfnamefont{V.}~\bibnamefont{Hernandez}},
  \bibinfo{author}{\bibfnamefont{J.~E.} \bibnamefont{Roman}}, \bibnamefont{and}
  \bibinfo{author}{\bibfnamefont{V.}~\bibnamefont{Vidal}},
  \bibinfo{journal}{Lect. Notes Comput. Sci.} \textbf{\bibinfo{volume}{2565}},
  \bibinfo{pages}{377} (\bibinfo{year}{2003}).

\bibitem[{\citenamefont{Roman et~al.}(2017)\citenamefont{Roman, Campos, Romero,
  and Tomas}}]{slepc-users-manual}
\bibinfo{author}{\bibfnamefont{J.~E.} \bibnamefont{Roman}},
  \bibinfo{author}{\bibfnamefont{C.}~\bibnamefont{Campos}},
  \bibinfo{author}{\bibfnamefont{E.}~\bibnamefont{Romero}}, \bibnamefont{and}
  \bibinfo{author}{\bibfnamefont{A.}~\bibnamefont{Tomas}}, \bibinfo{type}{Tech.
  Rep.} \bibinfo{number}{DSIC-II/24/02 - Revision 3.8},
  \bibinfo{institution}{D. Sistemes Inform\`atics i Computaci\'o, Universitat
  Polit\`ecnica de Val\`encia} (\bibinfo{year}{2017}).

\bibitem[{\citenamefont{Kelley}(2003)}]{Kelly}
\bibinfo{author}{\bibfnamefont{C.}~\bibnamefont{Kelley}},
  \emph{\bibinfo{title}{Solving Nonlinear Equations with Newton's Method}}
  (\bibinfo{publisher}{Society for Industrial and Applied Mathematics},
  \bibinfo{year}{2003}).

\bibitem[{\citenamefont{Balay et~al.}(2017{\natexlab{a}})\citenamefont{Balay,
  Abhyankar, Adams, Brown, Brune, Buschelman, Dalcin, Eijkhout, Gropp, Kaushik
  et~al.}}]{petsc-web-page}
\bibinfo{author}{\bibfnamefont{S.}~\bibnamefont{Balay}},
  \bibinfo{author}{\bibfnamefont{S.}~\bibnamefont{Abhyankar}},
  \bibinfo{author}{\bibfnamefont{M.~F.} \bibnamefont{Adams}},
  \bibinfo{author}{\bibfnamefont{J.}~\bibnamefont{Brown}},
  \bibinfo{author}{\bibfnamefont{P.}~\bibnamefont{Brune}},
  \bibinfo{author}{\bibfnamefont{K.}~\bibnamefont{Buschelman}},
  \bibinfo{author}{\bibfnamefont{L.}~\bibnamefont{Dalcin}},
  \bibinfo{author}{\bibfnamefont{V.}~\bibnamefont{Eijkhout}},
  \bibinfo{author}{\bibfnamefont{W.~D.} \bibnamefont{Gropp}},
  \bibinfo{author}{\bibfnamefont{D.}~\bibnamefont{Kaushik}},
  \bibnamefont{et~al.}, \emph{\bibinfo{title}{{PETS}c {W}eb page}},
%  \bibinfo{howpublished}{\url{http://www.mcs.anl.gov/petsc}}
  (\bibinfo{year}{2017}{\natexlab{a}}),
  \urlprefix\url{http://www.mcs.anl.gov/petsc}.

\bibitem[{\citenamefont{Balay et~al.}(2017{\natexlab{b}})\citenamefont{Balay,
  Abhyankar, Adams, Brown, Brune, Buschelman, Dalcin, Eijkhout, Gropp, Kaushik
  et~al.}}]{petsc-user-ref}
\bibinfo{author}{\bibfnamefont{S.}~\bibnamefont{Balay}},
  \bibinfo{author}{\bibfnamefont{S.}~\bibnamefont{Abhyankar}},
  \bibinfo{author}{\bibfnamefont{M.~F.} \bibnamefont{Adams}},
  \bibinfo{author}{\bibfnamefont{J.}~\bibnamefont{Brown}},
  \bibinfo{author}{\bibfnamefont{P.}~\bibnamefont{Brune}},
  \bibinfo{author}{\bibfnamefont{K.}~\bibnamefont{Buschelman}},
  \bibinfo{author}{\bibfnamefont{L.}~\bibnamefont{Dalcin}},
  \bibinfo{author}{\bibfnamefont{V.}~\bibnamefont{Eijkhout}},
  \bibinfo{author}{\bibfnamefont{W.~D.} \bibnamefont{Gropp}},
  \bibinfo{author}{\bibfnamefont{D.}~\bibnamefont{Kaushik}},
  \bibnamefont{et~al.}, \bibinfo{type}{Tech. Rep.} \bibinfo{number}{ANL-95/11 -
  Revision 3.8}, \bibinfo{institution}{Argonne National Laboratory}
  (\bibinfo{year}{2017}{\natexlab{b}}),
  \urlprefix\url{http://www.mcs.anl.gov/petsc}.

\bibitem[{\citenamefont{Balay et~al.}(1997)\citenamefont{Balay, Gropp, McInnes,
  and Smith}}]{petsc-efficient}
\bibinfo{author}{\bibfnamefont{S.}~\bibnamefont{Balay}},
  \bibinfo{author}{\bibfnamefont{W.~D.} \bibnamefont{Gropp}},
  \bibinfo{author}{\bibfnamefont{L.~C.} \bibnamefont{McInnes}},
  \bibnamefont{and} \bibinfo{author}{\bibfnamefont{B.~F.} \bibnamefont{Smith}},
  in \emph{\bibinfo{booktitle}{Modern Software Tools in Scientific Computing}},
  edited by \bibinfo{editor}{\bibfnamefont{E.}~\bibnamefont{Arge}},
  \bibinfo{editor}{\bibfnamefont{A.~M.} \bibnamefont{Bruaset}},
  \bibnamefont{and} \bibinfo{editor}{\bibfnamefont{H.~P.}
  \bibnamefont{Langtangen}} (\bibinfo{publisher}{Birkh{\"{a}}user Press},
  \bibinfo{year}{1997}), pp. \bibinfo{pages}{163--202}.

\bibitem[{\citenamefont{Theocharis et~al.}(2003)\citenamefont{Theocharis,
  Frantzeskakis, Kevrekidis, Malomed, and Kivshar}}]{gtheo}
\bibinfo{author}{\bibfnamefont{G.}~\bibnamefont{Theocharis}},
  \bibinfo{author}{\bibfnamefont{D.~J.} \bibnamefont{Frantzeskakis}},
  \bibinfo{author}{\bibfnamefont{P.~G.} \bibnamefont{Kevrekidis}},
  \bibinfo{author}{\bibfnamefont{B.~A.} \bibnamefont{Malomed}},
  \bibnamefont{and} \bibinfo{author}{\bibfnamefont{Y.~S.}
  \bibnamefont{Kivshar}}, \bibinfo{journal}{Phys. Rev. Lett.}
  \textbf{\bibinfo{volume}{90}}, \bibinfo{pages}{120403}
  (\bibinfo{year}{2003}).
%  \urlprefix\url{https://link.aps.org/doi/10.1103/PhysRevLett.90.120403}.

\bibitem[{\citenamefont{Kevrekidis et~al.}(2017)\citenamefont{Kevrekidis, Wang,
  Carretero-Gonz\'alez, and Frantzeskakis}}]{PhysRevLett.118.244101}
\bibinfo{author}{\bibfnamefont{P.~G.} \bibnamefont{Kevrekidis}},
  \bibinfo{author}{\bibfnamefont{W.}~\bibnamefont{Wang}},
  \bibinfo{author}{\bibfnamefont{R.}~\bibnamefont{Carretero-Gonz\'alez}},
  \bibnamefont{and} \bibinfo{author}{\bibfnamefont{D.~J.}
  \bibnamefont{Frantzeskakis}}, \bibinfo{journal}{Phys. Rev. Lett.}
  \textbf{\bibinfo{volume}{118}}, \bibinfo{pages}{244101}
  (\bibinfo{year}{2017}).
%  \urlprefix\url{https://link.aps.org/doi/10.1103/PhysRevLett.118.244101}.

\bibitem[{\citenamefont{Burger et~al.}(2002)\citenamefont{Burger, Carr,
  \"Ohberg, Sengstock, and Sanpera}}]{PhysRevA.65.043611}
\bibinfo{author}{\bibfnamefont{S.}~\bibnamefont{Burger}},
  \bibinfo{author}{\bibfnamefont{L.~D.} \bibnamefont{Carr}},
  \bibinfo{author}{\bibfnamefont{P.}~\bibnamefont{\"Ohberg}},
  \bibinfo{author}{\bibfnamefont{K.}~\bibnamefont{Sengstock}},
  \bibnamefont{and} \bibinfo{author}{\bibfnamefont{A.}~\bibnamefont{Sanpera}},
  \bibinfo{journal}{Phys. Rev. A} \textbf{\bibinfo{volume}{65}},
  \bibinfo{pages}{043611} (\bibinfo{year}{2002}).
%  \urlprefix\url{https://link.aps.org/doi/10.1103/PhysRevA.65.043611}.

\bibitem[{\citenamefont{Schulte et~al.}(2002)\citenamefont{Schulte, Santos,
  Sanpera, and Lewenstein}}]{PhysRevA.66.033602}
\bibinfo{author}{\bibfnamefont{T.}~\bibnamefont{Schulte}},
  \bibinfo{author}{\bibfnamefont{L.}~\bibnamefont{Santos}},
  \bibinfo{author}{\bibfnamefont{A.}~\bibnamefont{Sanpera}}, \bibnamefont{and}
  \bibinfo{author}{\bibfnamefont{M.}~\bibnamefont{Lewenstein}},
  \bibinfo{journal}{Phys. Rev. A} \textbf{\bibinfo{volume}{66}},
  \bibinfo{pages}{033602} (\bibinfo{year}{2002}).
%  \urlprefix\url{https://link.aps.org/doi/10.1103/PhysRevA.66.033602}.

\bibitem[{\citenamefont{Stellmer et~al.}(2008)\citenamefont{Stellmer, Becker,
  Soltan-Panahi, Richter, D\"orscher, Baumert, Kronj\"ager, Bongs, and
  Sengstock}}]{beckerr}
\bibinfo{author}{\bibfnamefont{S.}~\bibnamefont{Stellmer}},
  \bibinfo{author}{\bibfnamefont{C.}~\bibnamefont{Becker}},
  \bibinfo{author}{\bibfnamefont{P.}~\bibnamefont{Soltan-Panahi}},
  \bibinfo{author}{\bibfnamefont{E.-M.} \bibnamefont{Richter}},
  \bibinfo{author}{\bibfnamefont{S.}~\bibnamefont{D\"orscher}},
  \bibinfo{author}{\bibfnamefont{M.}~\bibnamefont{Baumert}},
  \bibinfo{author}{\bibfnamefont{J.}~\bibnamefont{Kronj\"ager}},
  \bibinfo{author}{\bibfnamefont{K.}~\bibnamefont{Bongs}}, \bibnamefont{and}
  \bibinfo{author}{\bibfnamefont{K.}~\bibnamefont{Sengstock}},
  \bibinfo{journal}{Phys. Rev. Lett.} \textbf{\bibinfo{volume}{101}},
  \bibinfo{pages}{120406} (\bibinfo{year}{2008}).
%  \urlprefix\url{https://link.aps.org/doi/10.1103/PhysRevLett.101.120406}.

\bibitem[{\citenamefont{Brand and Reinhardt}(2002)}]{brandas}
\bibinfo{author}{\bibfnamefont{J.}~\bibnamefont{Brand}} \bibnamefont{and}
  \bibinfo{author}{\bibfnamefont{W.~P.} \bibnamefont{Reinhardt}},
  \bibinfo{journal}{Phys. Rev. A} \textbf{\bibinfo{volume}{65}},
  \bibinfo{pages}{043612} (\bibinfo{year}{2002}).
%  \urlprefix\url{https://link.aps.org/doi/10.1103/PhysRevA.65.043612}.

\bibitem[{\citenamefont{Ma et~al.}(2016)\citenamefont{Ma, Navarro, and
  Carretero-Gonz\'alez}}]{Manjun}
\bibinfo{author}{\bibfnamefont{M.}~\bibnamefont{Ma}},
  \bibinfo{author}{\bibfnamefont{R.}~\bibnamefont{Navarro}}, \bibnamefont{and}
  \bibinfo{author}{\bibfnamefont{R.}~\bibnamefont{Carretero-Gonz\'alez}},
  \bibinfo{journal}{Phys. Rev. E} \textbf{\bibinfo{volume}{93}},
  \bibinfo{pages}{022202} (\bibinfo{year}{2016}).

\end{thebibliography}
\end{document}